\documentclass{pnastwo}
\usepackage{tabulary}

\url{www.pnas.org/cgi/doi/10.1073/pnas.0709640104}
\copyrightyear{2008}
\issuedate{Issue Date}
\volume{Volume}
\issuenumber{Issue Number}

\begin{document}
\title{Bursts of activity in collective cell migration}
\author{Oleksandr Chepizhko\affil{1}{Department of Applied Physics, 
		Aalto University, P.O. Box 11100, FIN-00076 Aalto, Espoo, Finland}, 
		Costanza Giampietro\affil{2}{Center for Complexity and Biosystems, Department of Biosciences,  University of Milan, via Celoria 26, 20133 Milano, Italy},
		Eleonora Mastrapasqua \affil{3}{Department of Biosciences, University of Milan, via Celoria 26, 20133 Milano, Italy},
		Mehdi Nourazar \affil{1}{},
		Miriam Ascagni\affil{3}{},
		Michela Sugni\affil{2}{},
		Umberto Fascio\affil{3}{},
		Livio Leggio\affil{3}{}, 
		Chiara Malinverno \affil{4}{Dipartimento di Scienze della Salute, San Paolo, University of Milan,
		20122 Milan, Italy}, 
		Giorgio Scita\affil{4}{},
		Stephane Santucci\affil{5}{Laboratoire de Physique de l\' {\'E}cole Normale Sup{\'e}rieure de Lyon,
		CNRS and Universit{\'e} de Lyon, 69364 Lyon, France}
		Mikko J. Alava\affil{1}{}
		Stefano Zapperi\affil{6}{Center for Complexity and Biosystems, 
		Department of Physics,  University of Milan, via Celoria 16, 20133 Milano, Italy}
	\affil{7}{Institute for Scientific Interchange Foundation, Via Alassio 11/C, 10126 Torino}
	\affil{8}{Istituto di Chimica della Materia Condensata e di Tecnologie per l'Energia, CNR-Consiglio Nazionale delle Ricerche, Via R. Cozzi 53, Milano 20125, Italy}\affil{1}{}, 
	Caterina A. M. La Porta\affil{2}{}}

\contributor{Submitted to Proceedings of the National Academy of Sciences
of the United States of America}

\significancetext{During wound healing and in cancer invasion cells migrate collectively driven by active internal forces and invade the available space. Here we show that 
this motion occurs by intermittent bursts of activity described by universal scaling
laws similar to the ones observed in other driven systems where a front
propagates in response to an external force, such as in fracture and fluid imbibition. Our results 
demonstrate that living systems display universal non-equilibrium critical fluctuations, induced by cell mutual interactions, that are usually associated to externally driven inanimate media.}
\maketitle

\begin{article}
\begin{abstract}
Dense monolayers of living cells display intriguing relaxation dynamics, reminiscent of soft and glassy materials close to the jamming transition, and migrate collectively
when space is available, as in wound healing or in cancer invasion. Here we show that  
collective cell migration occurs in bursts that are similar to those recorded in the propagation of cracks, fluid fronts in porous media and ferromagnetic domain walls. In analogy with these systems, the distribution of activity bursts displays scaling laws that are universal in different cell types
and for cells moving on different substrates. The main features of the invasion dynamics are quantitatively captured by a model of interacting active particles moving in a disordered landscape. Our results illustrate that collective motion of living cells is analogous to the corresponding dynamics in driven, but inanimate, systems.
\end{abstract}

\section*{Introduction}

Collective cell movement depends on intracellular biological mechanisms as well as 
environmental cues due to the extracellular matrix (ECM)\cite{Tambe2011,Brugues2014,Haeger2014,Lange2013,koch2012}, 
mainly composed by collagen which is organized in hierarchical structures,
such as fibrils and fibers. The mechanical properties of collagen fibril networks are essential 
to offer little resistance and high sensitivity to small deformations, allowing easy local remodeling 
and strong strain stiffening needed to ensure cell and tissue integrity \cite{Sacks2003}.
Wound healing is a typical biological assay to study collective migration of cells under
controlled conditions {\it in vitro} and is a prototypical experimental method to study active matter \cite{Vedula2013,Szabo2006,Poujade2007,Sepulveda2013}. Experiments performed on
soluble collagen \cite{Haga2005} or other gels \cite{Ng2012}, micro-patterned 
\cite{Saez2007,Rottgermann2014} and deformable substrates \cite{Tambe2011} show that 
cell migration is guided by the substrate structure and stiffness \cite{Oakes2009,koch2012,Metzner2015}. 

It has been argued that collective migration properties arise from stresses transmitted between neighboring cells \cite{Tambe2011} giving rise to long-ranged stress waves in the monolayer \cite{Serra-Picamal2012,Banerjee2015}. Hence the dynamics of an invading cell sheet is ruled by a combination of long-range internal stresses and interactions with the substrate, suggesting an analogy with driven elastic systems moving in a disordered medium such as cracks lines \cite{maloy2006,Tallakstad2011}, imbibition fronts \cite{Clotet2014} or ferromagnetic domain walls \cite{durin00}.  The scaling laws in these systems are usually associated with a depinning critical point that has been widely studied by simple models for interface dynamics. Thanks to a combination of numerical simulations \cite{leschhorn97,rosso2009} and renormalization group theory \cite{narayan92,leschhorn97,chauve01,ledoussal2009}, we now have a detailed picture of the non-equilibrium phase transitions and universality classes in these systems. 
Here we substantiate the analogy between collective cell migration and depinning by revealing and characterizing widely distributed bursts of activity in the collective migration of different types cells (human cancer cells and epithelial cells, mouse endothelial cells) over different substrates
(plastic, soluble and fibrillar collagen) and experimental conditions (VE-cadherin knock down) and compare  the experiments with simulations of a computational model of active particles \cite{Sepulveda2013}. We find that in all these cases the statistical properties of the bursts follow universal scaling laws that are quantitatively similar to those observed in driven disordered systems \cite{sethna01}. 
 

\section*{Results and discussion}
\subsection*{Cell front dynamics and activity clusters}
\begin{figure}[htb]\centering 
\includegraphics[width=\columnwidth]{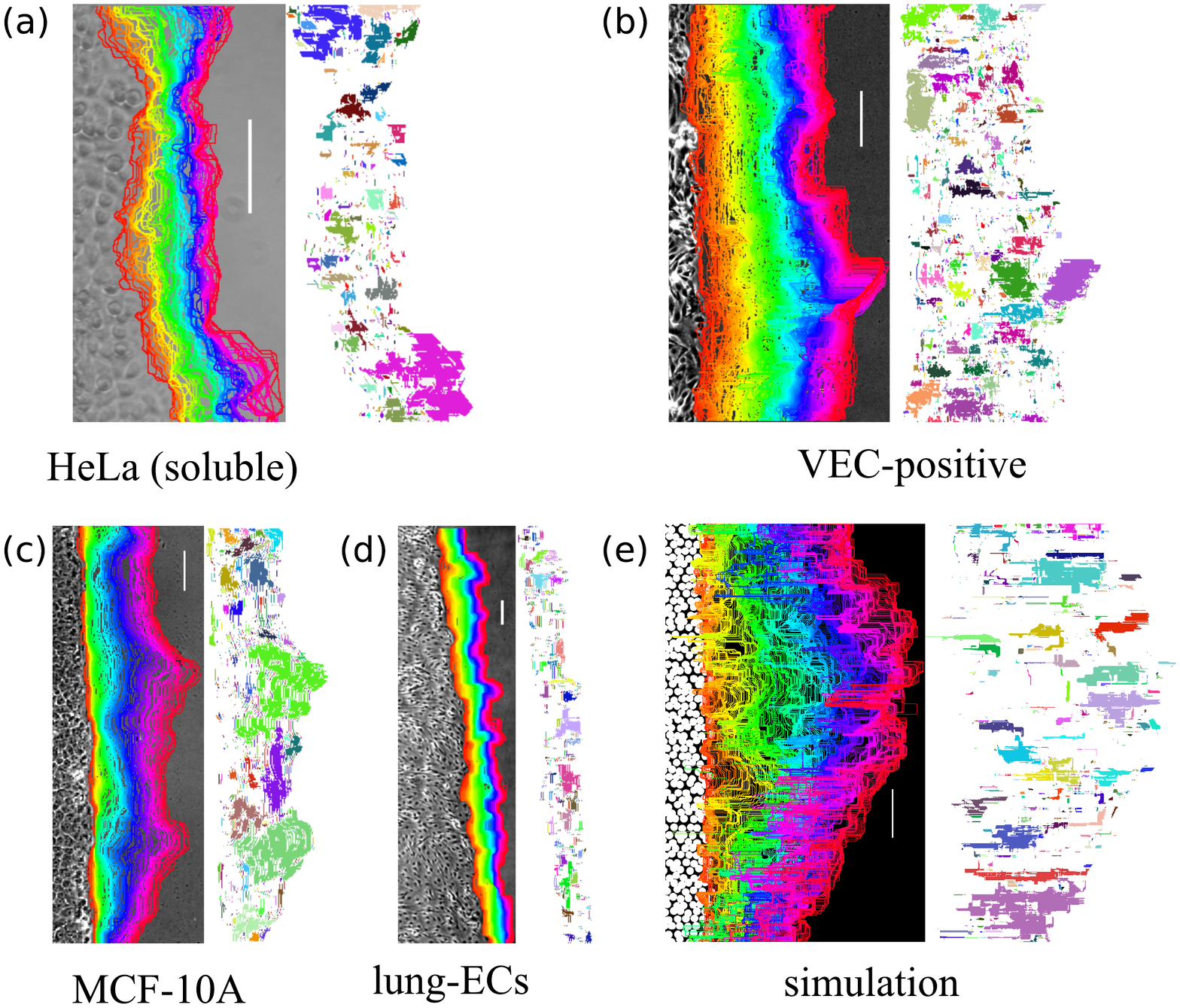}
 \caption{{\bf Cell front dynamics displays activity bursts.} Examples of cell fronts and their time evolution and the corresponding activity maps. Cell fronts are colored according to time. 
Regions marked by the same color in the activity map move collectively. The scale bars are 100$\mu$m.
a) HeLa cells moving on soluble collagen substrates. b) Mouse endothelial cells derived from embryonic stem cells (VEC-positive).  c) Human mammary epithelial cells  (MCF-10A).  d) Mouse endothelial cells extracted from lungs (lung-ECs).  e) Numerical simulations of the model.}
\label{fig:1} 
\end{figure}

We perform migration experiments on different cell lines as described in the Materials and
Methods section. We extract and record the cell front position as a function of
time as shown in Fig. \ref{fig:1} for experiments (see Videos S1-S7) and in simulations (Video S8). The fronts show  a rough structure with localized bursts of activity that is reminiscent of elastic lines moving in a pinning landscape \cite{maloy2006,Tallakstad2011,Clotet2014,durin00}. Spatio-temporal velocity fluctuations in moving fronts can be effectively quantified by constructing activity maps, as done previously for moving crack fronts \cite{Tallakstad2011} and fluid imbibition through porous media \cite{Clotet2014}.
As discussed in the Materials and Methods section, from each time-lapse movie 
we construct a spatial map $v^f(x,y)$ measuring the velocity of the front at position $(x,y)$
(see Fig. 
S1).   Using these maps we can define regions of coordinated activity 
by introducing a threshold $c$ for the velocity. The maps reported in Fig. \ref{fig:1} vividly illustrate the formation of localized clusters of activity similar to those found in fracture \cite{Tallakstad2011} and imbibition \cite{Clotet2014}. The corresponding distributions of cluster areas are reported in Fig. \ref{fig:2}a for experiments on different cell types and for simulations. They all display a power law decay $P(S) \sim S^{-\tau}$ up to a cutoff length $S^*$.  

The exponent $\tau$ of the power law distribution appears to be similar for different 
cell types: the fitted values (see Table S1) 
are all similar within error bars ($\tau\simeq 1.4-1.6$) and remarkably close to the values recorded in crack front propagation \cite{Tallakstad2011}.   As observed in moving cracks fronts, the cluster distributions, and in particular the cutoff to the power law decay, depends on the threshold $c$ used to identify the activity (see Materials and Methods and Fig. S1).
The value of the cutoff $S^*$ is in all cases much larger than the typical size of a cell indicating that correlated activity spans multiple cells (Fig. S2). 
We also provide additional evidence for universality by measuring the relation between width, length and size of clusters  Fig. \ref{fig:2}cd. Also in this case, we find robust scaling across different cell lines and experimental conditions.
 
\subsection*{Cell migration on collagen matrix}
To study quantitatively the effect of the substrate on the migration capability of cancer cells, we  study wound healing under three different cell coating conditions: plastic (i. e. no coating), bovine soluble and fibrillar collagen. The two types of collagen substrates are exactly the same in terms of biochemistry, but differ in the structural organization. In Fig. S3, 
we show the fibrillar organization of our collagen substrate by scanning electron microscopy (SEM), while soluble collagen substrates display no fibrils or structure. As shown in Fig.  \ref{fig:2}b, the exponent of 
the activity cluster distribution is independent on the substrate. The average cluster size, however, is larger for fibrillar collagen than for soluble collagen and plastic, independently on the value of $c$ (see Fig \ref{fig:2}b and Fig. S4a) 
To further quantify the role of the substrate in front propagation, we record  average position  of front as a function of time. As shown Fig. S5a, 
cell front motion is significantly slower  on plastic substrates than on collagen ones. There is no noticeable difference in the average velocity of fronts moving on soluble and fibrillar collagen substrates. Fluctuations do, however, differ as shown by the standard deviation of the front position that is larger for cells moving on fibrillar collagen substrates (Fig S5b) 

\subsection*{Velocity distributions}
\begin{figure}[htb] \centering 
\includegraphics[width=\columnwidth]{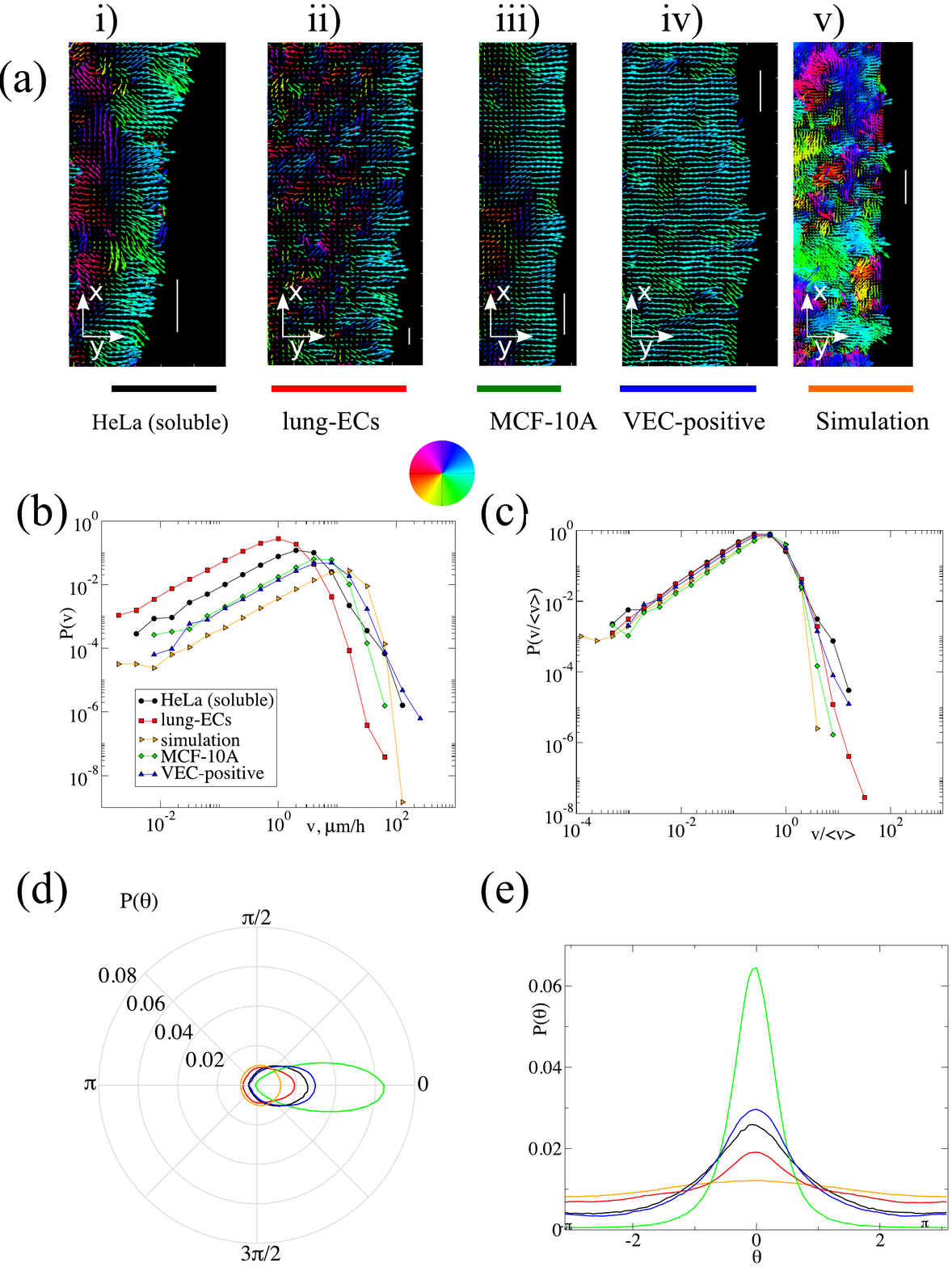}
 \caption{\label{fig:3} {\bf Cell velocity distributions display universal statistics} a) 
 Velocity maps for different cell types and for simulations as obtained from PIV. The colors indicate the orientation of the velocity following the color wheel. The length of the arrows is
 proportional to the magnitude of the velocity. The scale bars are 100$\mu$m.  b) The corresponding distributions of velocity magnitudes. c) Scaling the velocities by their magnitude leads to
 a collapse of all the distributions, apart from small deviations in the tails.
 d,e) The distribution of the orientations of the velocities.} 
 \end{figure}

To better quantify the properties of the different experimental conditions,
we resort to particle image velocimetry (PIV, see Materials and Methods) which allows
to estimate local velocities treating the cell layer as a fluid. Representative examples of the local 
velocities for different cell types and for simulations are reported in Fig. \ref{fig:3}a
which displays the orientation of the velocity field (see also Videos S1-S7). While cells mostly advance towards the empty space there is significant motion also in the transverse and even in the opposite direction. This is summarized in the orientation distribution, revealing small differences due
to cell types and substrates (Fig. \ref{fig:3}de and S5c). 
Next, in Fig. \ref{fig:3}b and S5d 
we report the distribution of the absolute value of the velocities extracted from PIV, indicating again quantitative differences between
different cell types and substrates. Yet, the general shape of the velocity distribution 
is similar in all cases as shown in Fig. \ref{fig:3}c where the velocities for each experiment
are rescaled by their average value. This leads to a collapse of all the distributions, including the results of numerical simulations, apart from small deviations in the tails.

\subsection*{Effect of VE-cadherin knockdown}
\begin{figure}[htb] \centering 
\includegraphics[width=\columnwidth]{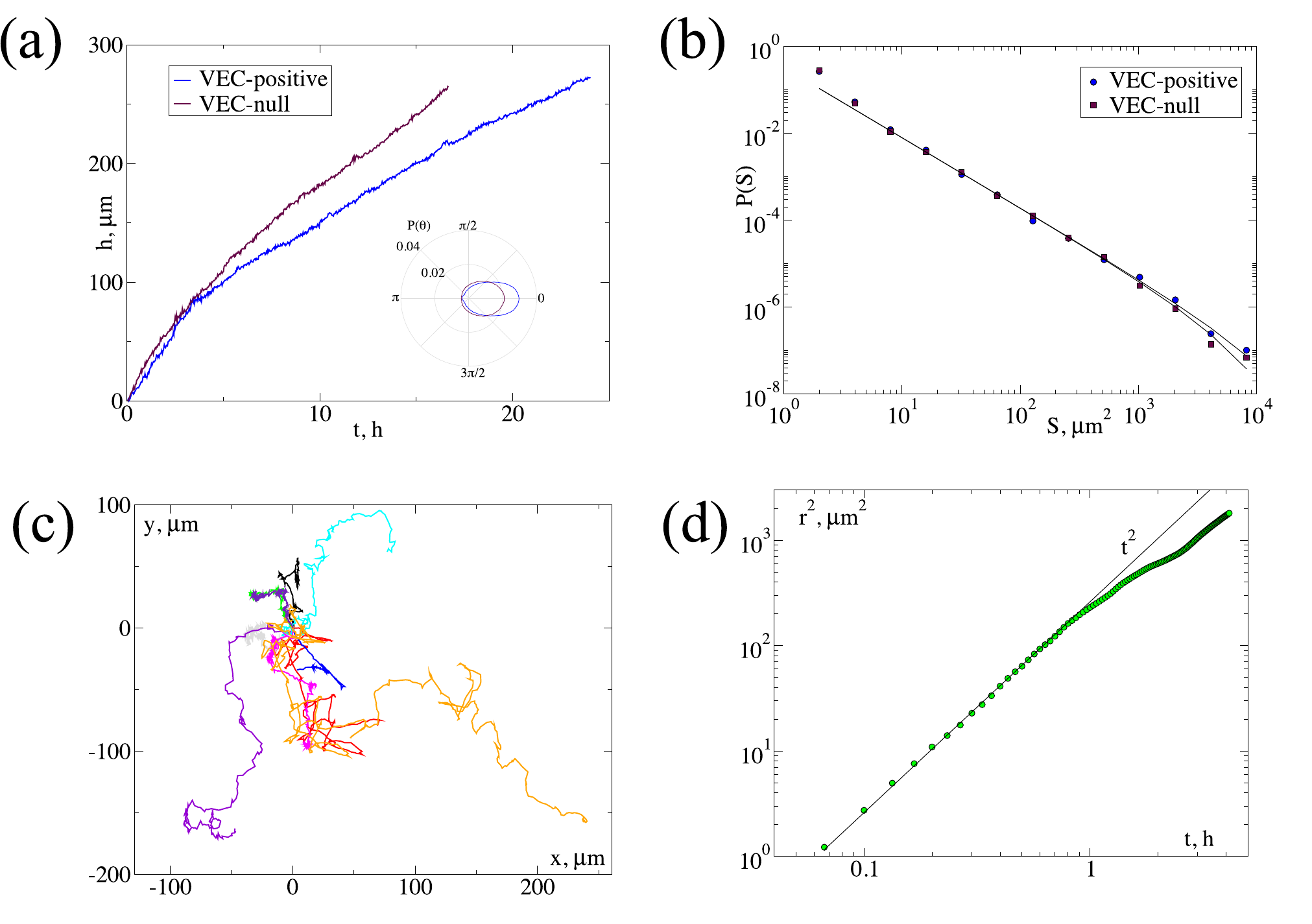}
 \caption{\label{fig:4} {\bf Knock down of VE-cadherin leads to faster fronts and individual
 cell motion}. a) The time evolution of the front position for mouse endothelial cell expressing
 (VEC-positive) or not (VEC-null) VE-cadherin. VEC-null cells move faster. b) The corresponding 
 cluster size distribution displays the same exponent and small changes in the cutoff. c) 
 In the VEC null cases cells detach from the front and invade the space individually following
 trajectories as the ones illustrated. d) The average mean-square displacement of the trajectories
 indicates an initial ballistic regime followed by a slowing down.  }
 \end{figure}

Internal couplings usually play a key role in avalanche statistics \cite{narayan92,leschhorn97,chauve01,ledoussal2009}.
To understand this point in our context, we assess the role of cell-cell interactions performing the same
cell migration assay knocking down VE-cadherin in mouse endothelial cells \cite{giampietro2015}.
VE-cadherin is an endothelial specific cadherin molecule located at adherent junctions which regulats adhesion between adjacent endothelial cells. Here we consider cells without VE-cadherin (VEC-null) but still expressing N-cadherin, another 
important molecule expressed by several types of cells such as neuronal, skeletal and heart muscle cells or fibroblasts \cite{Goodwin2004}. Moreover, N-cadherin is highly expressed in mesenchymal stem cells and cancer cells that are highly invasive and poorly polarized \cite{Wheelock2003}. As shown in Fig. \ref{fig:4}a, VEC-null cell fronts move
faster, but in a more disordered way (see Fig. S6),
than VEC-positive fronts. 
The distribution of activity cluster areas is very similar in the two cases (Fig. \ref{fig:4}a), 
with slightly larger cutoff size for VEC-positive cells (see Fig. S2).
Despite these similarities,
VEC-null cells tend to detach from the advancing front and move forward (see Video S9). We have tracked the
trajectories of individual cells (Fig. \ref{fig:4}c) and shown that they follow a persistent random walk, displaying
ballistic motion at short time-scales (Fig. \ref{fig:4}d). The combination of faster fronts and individual
cells detachment confirms a more invasive phenotype for VEC-null cells \cite{giampietro2012}.

\subsection*{Numerical simulations of an active particle model}
In order to better understand how avalanche-like fluctuations arise in collective cell migration,
we simulate the model of interacting active particle introduced in 
Ref. \cite{Sepulveda2013} (for the details see the Materials and Methods section). 
The geometry of the system, the size and number of particles are chosen to match the experimental images. Particles are first packed into a confined space and then allowed to invade space filled with ``surface'' particles. For the time scales considered in the present experiments, cell division is negligible and therefore we do not consider it in the model. The main tuning parameters of the model are the inter-particle velocity coupling $\beta$, the amplitude of the noise $\eta$ and the adhesion strength $U_1$. We also take into account the rough structure of the fibrillar collagen substrate by including a quenched (i.e. time independent) random force field with amplitude $\sigma_{\mathrm rf}$ and correlation length $\zeta$.  Simulations representing plastic or soluble collagen substrate are performed without quenched disorder (i.e. $\sigma_{\mathrm rf}=0$), with fluctuations arising only from the random initial condition and the  time-dependent noise in the dynamics. In addition to the random force field, the main difference between our simulations and Ref. \cite{Sepulveda2013} is that we do not consider leader cells. 

\begin{figure}[htb] \centering 
\includegraphics[width=\columnwidth]{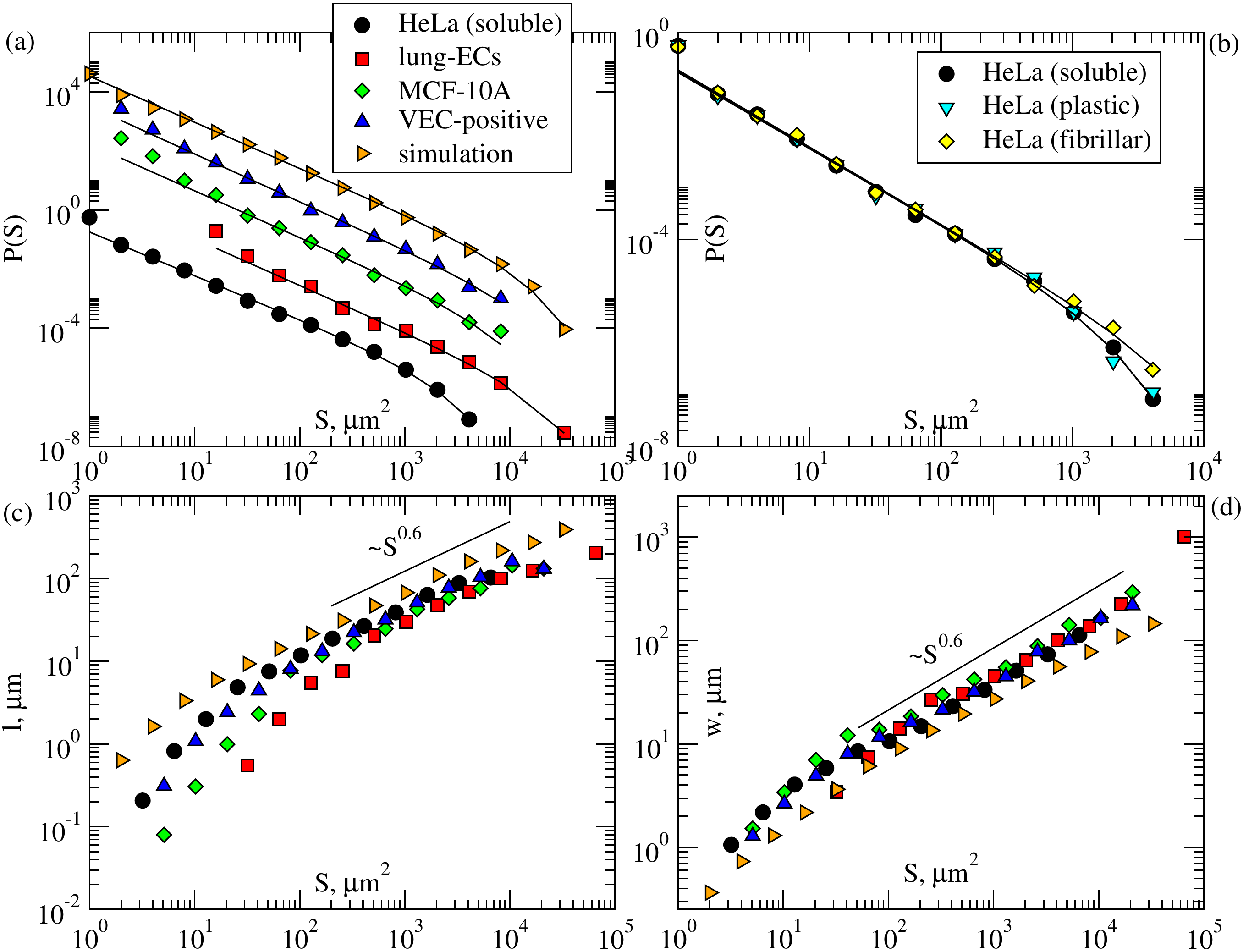}
 \caption{\label{fig:2} {\bf Cell activity bursts statistics displays scaling} a) Distributions of the areas of activity clusters display power law scaling with a cutoff. The distributions for different cell types have been shifted for clarity. The slope obtained fitting the distributions is very similar for all cell types. b) Distributions of the areas of activity clusters for HeLa cells moving on different substrates. The scaling exponent is the same but the cutoff size for fibrillar collagen substrates is larger than for plastic and soluble collagen substrates. c) The average cluster length and d) the average cluster width scale with the cluster size with an exponent that is independent on the cell type.} 
 \end{figure}

The numerical simulations allow to reproduce with good accuracy all the salient features of collective cell migration revealed by our experiments as reported in Figs. \ref{fig:1}-\ref{fig:3}. As already discussed, the numerical simulations reproduce the statistical features of activity bursts (Fig. \ref{fig:2}) and cell velocities (Fig. \ref{fig:3}). Furthermore, the numerical results illustrate the key role played by the substrate disorder, encoded in its correlation length $\zeta$, in determining the cutoff scale of the cluster size distribution (see Fig. S7).
In particular, higher values of $\zeta$ correspond to larger activity clusters. This explains why cells move in larger activity bursts on fibrillar substrates, where disorder is stronger. Decreasing the velocity coupling $\beta$ or the 
adhesion strength $U_1$ increases the front velocity (Fig. S8) 
which should be compared with an analogous result observed when VE-cadherin is knocked down (Fig. \ref{fig:4}).  The model allows to explore the limits of universality. We observe an activity cluster distribution comparable to the experimentally measured one only for relatively small values of $\beta$. When $\beta$ is larger (e.g. $\beta = 60$) the exponent $\tau$ is significantly smaller, suggesting the presence of different universality classes (see Fig. S7
and Table S2
for the results of the fits). Finally, we assess the relevance of interactions
(i.e. performing simulations with $\beta=0$ and $U_0=U_1=0$) and pinning
(i.e. no surface particles and no random field) and find that 
when either one of the two ingredients is missing the distribution is not a power law anymore (see Fig. S9). When neither pinning and interactions are present, we can not record a well defined front.  

\subsection*{Universal scaling}
\begin{figure}[htb] \centering 
\includegraphics[width=\columnwidth]{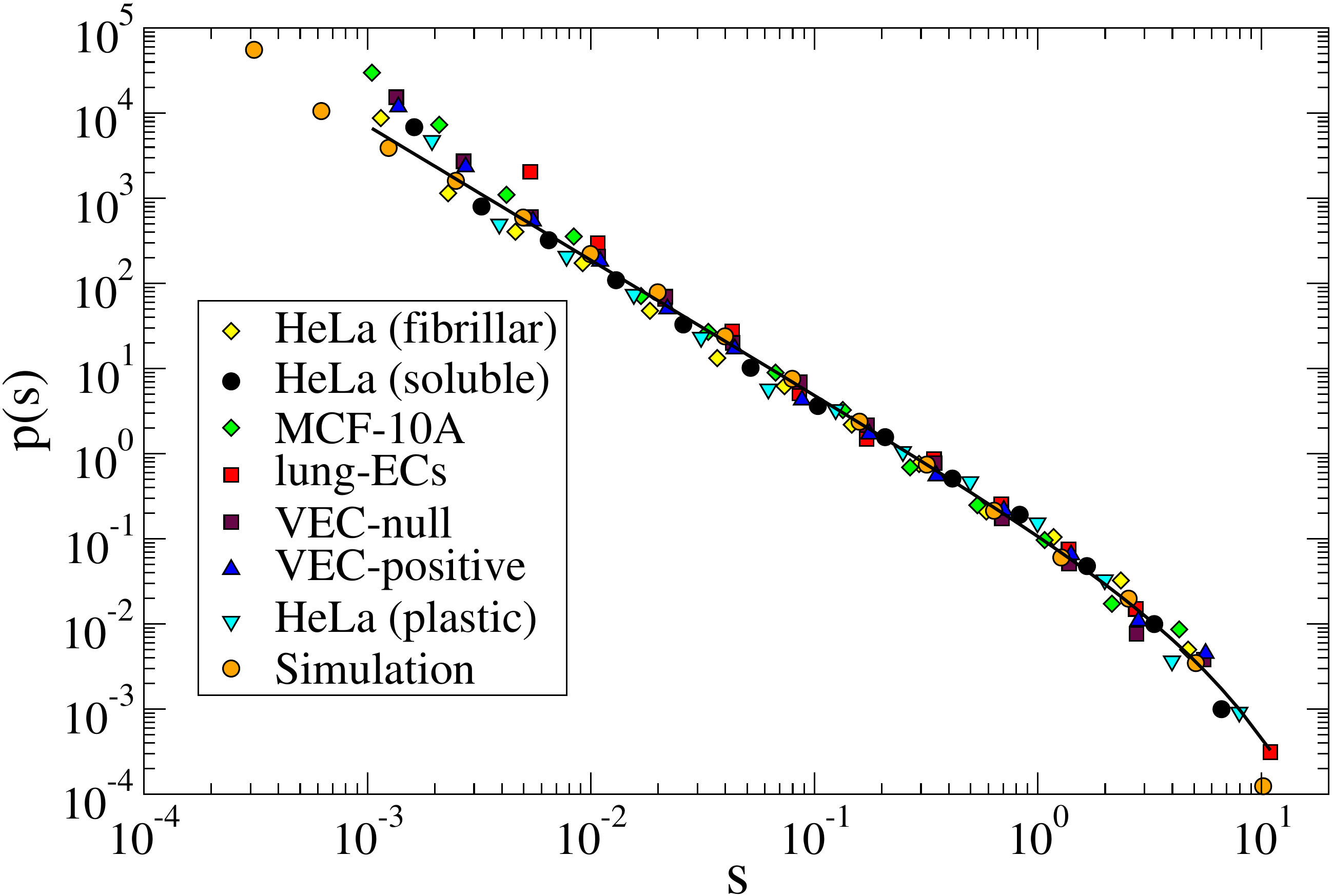}
 \caption{\label{fig:5} {\bf The distribution of activity clusters is universal}. The distributions of 
 the areas of activity clusters from Figs. \ref{fig:2} and \ref{fig:4} can all be collapsed into
 a single universal scaling function $p(s)=s^{-\tau}\exp(-C s)$ (line) when plotted in terms of the
  reduced variable $s=S/S^*$. }
 \end{figure}

The combinations of experimental results and numerical simulations suggests that the 
activity bursts observed in collective cell migration could be described by universal scaling
laws as in non-equilibrium critical phenomena. In order to strengthen this conclusion it is, however,
imperative to overcome the limitations of power law fitting \cite{clauset2009} and focus instead on scaling
functions, as indicated by Sethna et al. \cite{sethna01}. Ref. \cite{ledoussal2009,rosso2009} shows that in proximity of a depinning critical point, avalanche distributions $P(S)$ can be written as 
$P(S)=\frac{\langle S\rangle}{S^*} p(S/S^*)$, where  $S^* = \frac{\langle S^2 \rangle}{2\langle S \rangle}$ and $p(s)$ is a universal scaling function. Here, we follow the same approach
with our data and show that indeed all the experimental distributions, involving different
cell lines and substrate, and those obtained from numerical simulations collapse into
a single universal scaling curve as shown in Fig. \ref{fig:5}. We thus perform a joint fit
of all the distributions with a single scaling function $p(s)=s^{-\tau} \exp(-C s)$, yielding
$\tau=1.58\pm 0.02$ as best fitting parameter. This result provides a strong test of the
universality of the bursty behavior we observe in collective cell migration.

\subsection*{Discussion} 
The collective motion of a cell layer as it invades an empty region has been extensively studied experimentally \cite{Vedula2013,Szabo2006,Poujade2007,Sepulveda2013} but the relation with the glassy features observed in confluent layers \cite{Angelini2011,Park2015} was not explored. The externally driven motion of disordered and glassy systems typically involves intermittent behavior and avalanches \cite{sethna01}. Our work shows that collective bursts of activity are also present in active matter with cells advancing in avalanches of widely distributed sizes. Similar intermittent activity \cite{Weber2015} and scaling behavior has been previously recorded in animal \cite{Ginelli2015} and even human mobility patterns \cite{Song2010} and it is thus intriguing to realize that it exists  even at the cellular level.  Finally, migration is a key property of tumor cells for tissue invasion and metastasis.
Our detailed statistical analysis shows quantitatively how the organization
and structure of the substrate, either gel or fibrillar, affects the way cells move. The final invasion
velocity is the same but the motion is different, with larger intermittent fluctuations present on fibrillar substrates.  This finding is interesting since it implies that cancer cells use different internal mechanisms to migrate depending on the environment, a fact that should be relevant for the metastatic process.

\section*{Materials and Methods}
\subsection*{Collagen coating}
Collagen substrates are prepared from a 3\% solution of soluble bovine collagen in 0.01 HCl (code C4243, Sigma), kept in ice until use. Both solubilized ("soluble collagen") and re-fibrillated collagen ("fibrillar collagen") are prepared in 35 mm Petri dishes. For soluble substrates a sufficient amount of collagen solution is added to a Petri dish so that the bottom is covered, the excess is removed after 15 minutes and the dish is dried at room temperature for at least 2 hours. For fibrillar substrates, a mixture of 3\% collagen solution (Code C4243, Sigma), 0.1N NaOH, 10\% PBS (8:1:1) is prepared and 0.8mL are pipetted to the petri dish and incubated at 37$^\circ$ C overnight to allow the proper gelification/refibrillation. After polymerization, samples are 5-10 micrometers in height.

\subsection*{Cell culture}
HeLa cell line (ATCC CCL-2) is cultured in DMEM medium supplemented with 10\% fetal calf serum, 2 mM L-glutamine, 100 U/ml penicillin, 100 mg/ml streptomycin and 0,25 mg/ml amphotericin B (Invitrogen, Milan, Italy) at 37$^\circ$C and humidity (95\% relative humidity), and CO2 concentration (5\%). 
Cells are seeded in bovine soluble collagen, or fibrillar bovine collagen or without collagen-coated dishes and growth up to reach confluence overnight. 
Endothelial cells derived from embryonic stem cells with homozygous null
mutation of the VE-cadherin gene (VEC null) \cite{balconi2000}. The
wild type form of VE-cadherin was introduced in these cells (VEC positive) as
described in detail in \cite{Lampugnani2002}  Endothelial cells isolated from lungs of wild type adult mice and cultured as previously described \cite{giampietro2015}.  Starving medium was MCDB 131
(Invitrogen) with 1\% BSA (EuroClone), 2-mM glutamine, 100 U/liter
penicillin/streptomycin, and 1-mM sodium pyruvate.
The MCF-10A cell line is a non-tumorigenic human mammary epithelial cell
line (ATCC CRL10317) grown in DMEM/F12 supplemented with 5\%horse serum,
20ng/ml EGF, 0.5mg/ml Hydrocortisone, 100ng/ml Choilera toxin, 10$\mu$g/ml
Insulin.

\subsection*{Migration assay}
For the migration assay, a wound is introduced in the central area of the confluent cell sheet by using a pipette tip and the migration followed by time-lapse imaging.  Hela cells were stained with 10 $\mu$M Cell tracker green CMFDA (Molecular Probes) in serum-free medium for 30 minutes and then the complete medium was replaced. Mouse endothelial cell monolayers were wounded after an overnight starving, washed with PBS, and incubated at 37$^\circ$C in starving medium. MCF-10A cell monolayers were wounded after an O/N doxycicline induction, washed with PBS, and incubated at 37$^\circ$C in fresh media+doxycicline.

\subsection*{Time lapse imaging}
For Hela cells, time-lapse multifield experiments were performed using an automated inverted Zeiss Axiovert S100 TV2 microscope (Carl Zeiss Microimaging Inc., Thornwood, NY) with a chilled Hamamatsu  CCD camera OrcaII-ER. Displacements of the sample and the image acquisition  are computer-controlled using Oko-Vision software (from Oko-lab).  This Microscope was equipped with a cage incubator designed to maintain all the required environmental conditions for cell culture all around the microscopy workstation, thus enabling to carry out prolonged observations on biological specimens. Cell Tracker and Phase contrast images were acquired with an A-Plan 10x (NA 0.25) objective; the typical delay between two successive images of the same field was set to 10 minutes for 12 hours. 
For mouse endothelial cells and MCF-10A cells, time-lapse imaging of cell migration was performed on an inverted microscope (Eclipse TE2000-E; Nikon) equipped with an incubation chamber (OKOLab) maintained at 37$^\circ$C in an atmosphere of 5\% CO2. Movies were acquired with a Cascade II 512 (Photometrics) charge-coupled device (CCD) camera controlled by MetaMorph Software (Universal Imaging) using a 4X or 10× magnification objective lens (Plan Fluor 10×, NA 0.30). Images were acquired every 2 or 5 min over a 24h period. See Table S3 
for the time lapse parameters.

 \subsection*{Scanning Electron Microscopy}
Collagen substrates are fixed with 2\% glutaraldehyde in 0.1 M cacodylate buffer  for 2 h at 4°C and post-fixed with 1\% osmium tetroxyde in 0.1 M sodium cacodylate buffer (2 h, room temperature). After several washings with dH2O, they are dehydrated in an increasing ethanol scale and treated with a series of solutions of HMDS (Hexamethyldisilazane) and ethanol in different proportions (1:3, 1:1, 3:1 and 100\% HMDS), mounted on stubs, covered with pure gold (Agar SEM Auto Sputter, Stansted, UK). The samples are observed under a scanning electron microscope(SEM) (LEO-1430, Zeiss, Oberkochen, Germany).

\subsection*{Particle image velocimetry (PIV)}
The measurements of the velocity field were done using PIVlab app for Matlab \cite{PIV0, PIV1}. The method is based on the comparison of the intensity fields of two consequent photographs of cells. The difference in the intensity is converted into velocity field measured in $px/frame$. Then the velocity is converted to $\mu m/h$ through coefficient that is specific for each experiment, shown in table S3. 

\subsection*{Border progression measurement}
The method of border detection is based on the procedures described in
~\cite{Johnston2014,Treloar2013}. The method is based on the border extraction procedure described in the manual of the Matlab software package (\texttt{CellSegmentationExample}). 

\subsection*{Front activity maps}
We analyze the propagation of the cell front, by measuring for each recorded image, and thus, at each time step $t$, the cell front position $y=h(x,t)$, along the abscissa $x$. We then construct a {\it local velocity map} of such interface $v^f(x,y)$, by computing the  velocity of the cell front at each position of the front ($x$, $y = h(x,t)$) during its progression:
\begin{equation*}
v^f(x,y=h(x,t_i))=\frac{h(x, t_{i+1}) - h(x, t_i)}{t_{i+1}-t_i}
\label{eq:vel_in_x_h}
\end{equation*}
Figure~S1a provides such spatial color scale map $v^f(x, h(x,t))$ of the local velocity fluctuations for a typical wound healing experiment using HeLa cells moving on a soluble collagen substrate. The various regions of different color levels reveal an intermittent burst-like dynamics on a broad range of length scales. Such a complex dynamics can also be unveiled by representing the spatio-temporal map $v^f(x,t)$ giving the instantaneous velocity of each point of the cell front during its propagation, as shown in Fig.~S1b.  Occasional overhangs in the front are eliminated by the maximum value of $h$ for a given $x$. When the front locally moves backwards, we average all the velocity values at that point. To define areas of correlated activity, we consider the spatial map of the local front velocities  $v^f(x, h(x,t))$ and define avalanches as clusters of velocities larger than an arbitrary threshold \cite{Tallakstad2011,Clotet2014}: $v^f(x, h(x,t)) \geq c \langle v^f \rangle$, where $\langle v^f \rangle$ is the mean front velocity during the experiment  Fig.~S1c. Cluster sizes $S$ are given by the area of the clusters,  and their shape is characterized by their length $l$ (lateral extension along $x$) and their width $w$ (extension in the mean direction of propagation $y$). The cutoff of the cluster size distribution depends slightly on $c$, but its scaling exponent does
not  (Fig.~S1d).

\subsection*{Simulations of the active particle model}
We simulate collective cell migration using a modified version of the model introduced in Ref. \cite{Sepulveda2013}.
Here we introduce a pinning field and we do not consider a leader cell.

The main equation of motion is given by
\begin{equation}
	\label{eq:main_Sepulveda}
	\frac{d \mathbf{v}_i}{d t} =
        - \alpha \mathbf{v}_i + \sum_{j}
        \left[
          \frac{\beta}{N_i}(\mathbf{v}_j-\mathbf{v}_i)+\mathbf{f}_{ij}
          \right]
        +
        \sigma(\rho_i) \boldsymbol{\eta}_i + \mathbf{F}_{\mathrm{rf}}(\boldsymbol{x}_i)\,
\end{equation}
where the sum is restricted to the nearest neighbors of $i$,
$\alpha$ is a damping parameter, $\beta$ is the velocity coupling strength, $\mathbf{f}_{ij}=-\nabla_i U(r_{ij})$ is 
the force between neighboring cells where
\begin{equation}
  \label{eq:potential}
  U(r)=U_0 \exp(-(r/a_0)^2)+U_1 (r-a_1)^2 H(r-a_1)\,.
\end{equation}
Here $H(x)$ is the Heaviside function $H(x)=1$ for $x>0$ and $H(x)=0$ otherwise, $U_1$ is the adhesion 
and $U_0$ the repulsive strength.
The noise term $\sigma(\rho_i) \boldsymbol{\eta}_i$, here $\boldsymbol{\eta}_i$ is an Ornstein-Uhlenbeck process with correlation time $\tau$:
\begin{equation}
  \label{eq:OUprocess}
  \tau \frac{d \boldsymbol{\eta}_i}{dt} = - \boldsymbol{\eta}_i + \boldsymbol{\xi}_i,
\end{equation}
$\boldsymbol{\xi}_i$ is a delta-correlated white noise, independent for each cell $\langle \boldsymbol{\xi}_i (t) \boldsymbol{\xi}_j (t') \rangle = \delta_{ij} \delta(t-t')$
The amplitude of the noise term $\sigma$ depends on the density of the neighboring particles $\rho_i$:
\begin{equation}
  \label{eq:sigma}
  \sigma (\rho_i) = \sigma_0+(\sigma_1-\sigma_0)(1-\rho_i/\rho_0).
\end{equation}
The neighbors of each particle are defined in the same way as Ref. \cite{Sepulveda2013} The neighborhood of a particle $i$ is split into $6$ equal sectors. The particle closest to the particle $i$ in each sector (but closer then interaction radius of $100$ $\mu m$) is chosen to be a neighbor. The local density is computed then as $\rho_i=1/[\pi (\langle d  \rangle/2)^2]$, where $\langle d  \rangle$ is the average distance between cell $i$ and its $6$ neighbors (for sectors without neighbors the distance is taken to be $100$ $\mu m$). 
To model heterogeneous substrates we include an additional random force field $\mathbf{F}_{\mathrm{rf}}(\mathbf{x}_i)$. A pair of Gaussian distributed random numbers, representing the two components of the random force, are placed on a square grid of spacing $\zeta$, quantifying the correlation length of the disorder. The value of random force between grid points is obtained through bilinear interpolation.

As in the original model, the free surface is modeled by surface particles, that are hindering cells to enter the empty space. The interaction between a surface particle and a cell is modeled as
\begin{equation}
  \label{eq:fs}
  \mathbf{f}_{ij}^s=-\nabla_i U^s(r_{ij})\,,
\end{equation}
where $U^s(r)$ is $U^s(r)=A_s \exp(-(r/a_s)^2)$.
A scalar damage variable $q$ is associated with each surface particles and follows the equation
\begin{equation}
  \label{eq:dtheta}
  \nu \frac{d q_i}{dt}=\sum_{j: \mathbf{r_j} \in S_i} |\mathbf{f_{ij}^s}|\,.
\end{equation}
The values of the parameters that were used for simulations match the ones 
used in Ref. \cite{Sepulveda2013}: number of particles $N=4000$, size of the box $2L \times L$, $L=1000$ $\mu\mathrm{m}$,
$\alpha=1.42\mathrm{h}^{-1}$, $\beta$, varying, but the value used in most of simulations was $\beta=10$ $\mathrm{h}^{-1}$, $\tau=1.39$ $\mathrm{h}^{-1}$, $\sigma_0=150$ $\mu \mathrm{m}\mathrm{h}^{-3/2}$, $\sigma_1=300$ $\mu \mathrm{m}\mathrm{h}^{-3/2}$, $U_0=2400$ $\mu \mathrm{m}^2/\mathrm{h}^2$, $a_0=8$ $\mu \mathrm{m}$, $U_1=2 \mathrm{h}^{-2}$, $a_1=35 \mu \mathrm{m}$, $\rho_0=4.0 \times 10^{-3}
\mu \mathrm{m}^{-2}$. $r_s$ the radius of interaction between a cell and a surface particle, is chosen
to be $r_s=50\mu\mathrm{m}$, $\nu=7 \mathrm{h}^{-1}$, $A_s=2400\mu \mathrm{m}^2/\mathrm{h}^2$, $a_s=8$ $\mu \mathrm{m}$, the threshold value $q_i=\theta$ after reaching which the surface particle disappears is taken to be $\theta=30$ $\mu \mathrm{m}$. The values for the parameters of the random field are also varied. Contrary to Ref. \cite{Sepulveda2013}, we do not introduce any leader cell.

\subsection*{Data fitting}
We fit the distribution of cluster sizes using the least-square method for binned histograms and the
maximum likelihood method (see Fig. S10, Tables S1, S2 and Supporting Information for more details).

\section*{Acknowledgments}
OC and SZ acknowledges support from the Academy of Finland FiDiPro program, project 13282993. CAMLP and SS thank the visiting professor program of Aalto University where part of this work was completed. SZ is supported by ERC Advanced Grant n. 291002 SIZEFFECTS.  MN, MJA  are supported by the Academy of Finland through its Centres of Excellence Programme (2012-2017) under project n. 251748. We acknowledge the computational resources provided by the Aalto University School of Science ``Science-IT'' project, as well as those provided by CSC (Finland). GS is supported by grants from the Associazione Italiana per la Ricerca sul Cancro (AIRC) (n. 10168),  Worldwide Cancer Research (AICR-14-0335), and the European
Research Council (Advanced ERC n. 268836) 

\section*{Authors Contributions}
OC, MN, SS analyzed the data. CAMLP, CG, LL, MA, MS, UF, CM, GS, EM performed the experiments. OC, MN performed numerical simulations. CAMLP designed the experiments. OC, MJA, SZ designed the model. CAMLP, SZ wrote the manuscript. CAMLP, MJA, SZ coordinated the project.


\newpage
 
\clearpage


\section*{Supplementary information}



\subsection*{Details about the fitting methods}
We employ different strategies to obtain the best parameters describing the cluster size distributions in simulations and experiments. We first consider the log-spaced binned probability density functions and perform a lest-square fitting
with the function   
\begin{equation}
\label{eq:powerlawcutoff}
f(x)=A x^{-\tau} \exp(-C x)\,,
\end{equation}
To obtain a reliable estimate of the cutoff, we first take the logarithm of the distribution and fit it with the
logarithm of Eq. \ref{eq:powerlawcutoff}. The fitting is performed in python using the pyFitting function (https://github.com/gdurin/pyFitting) and the results are reported in table \ref{table:S1} and \ref{table:S2}
for experiments and simulations, respectively.

The second strategy relies on the maximum-likelihood estimate. Given the set of 
measured cluster sizes $S_i$, we consider the log-likelihood function
\begin{equation}
\mathcal{L}= -\sum_i \log(f(S_i))),
\end{equation}
where the function $f(x)$ is given by Eq. \ref{eq:powerlawcutoff}. The function $\mathcal{L}$ is 
maximized with respect to the parameters $A$, $\tau$ and $\alpha$. The results are again
reported in table \ref{table:S1} and \ref{table:S2}. A comparison of the fitted
function and the data can be presented in a way that is independent on binning by plotting 
the cumulative distribution functions in Fig. \ref{fig:S10}. The theoretical function in this
case is given by
\begin{equation}
CDF=\label{eq:cdf}
\frac{\Gamma(1-\tau, C x)}
{\Gamma(1-\tau, C x_{min})}\,
\end{equation}
where $\Gamma(a,b)$ is the incomplete gamma function.

Finally, we collapse together all the experimental probability density function into a unique set. 
To this end, we first compute the characteristic cluster size
$S^* = \frac{\langle S^2 \rangle}{2\langle S \rangle}$ from each experimental data set. We then
rescale the cluster sizes, defining a reduced size $s=S/S^*$ and a scaling function  
$p(s)= P(S)\frac{S^*}{\langle S\rangle}$. Notice that $p(s)$ is not a distribution, since its integral
is not equal to one. Once data for different experiments are rescaled according to this prescription,
they are joined into a single set which is fitted by the least-square method using the function reported
in Eq. \ref{eq:powerlawcutoff}. The result yields $A=0.13\pm 0.01$, $\tau=1.58\pm 0.02$ and $C=0.20\pm 0.03$.

\pagebreak


\setcounter{figure}{0}
\renewcommand{\thefigure}{S\arabic{figure}} 

\begin{figure}[htb] \centering 
\includegraphics[width=8cm]{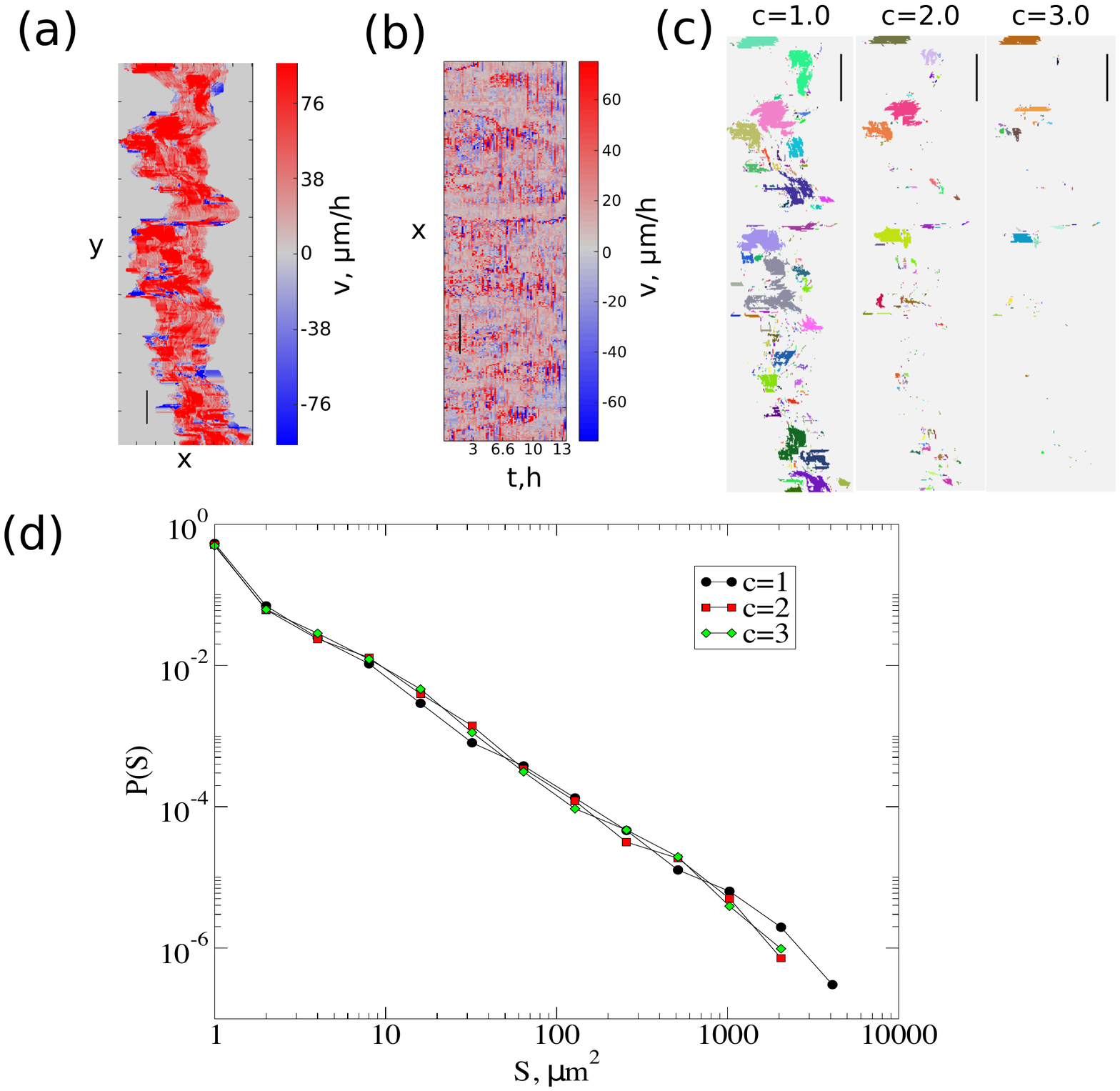}
 \caption{\label{fig:S1} Determination of activity areas from images. Scale bars are 100$\mu$m.
 a) Spatial front velocity maps. Each pixel is colored according to the velocity of the
 front once it passes through that location. b) Spatio-temporal velocity map. Each pixel corresponds
 to the velocity ($v_f(x,t)$) at time $t$ of the front at point $x$ c) Activity clusters are
 obtained by setting threshold $c$ to the data in panel a. Results obtained with different values
 of $c$ are reported in three panels. d) Cluster size distributions for different values of $c$.}
 \end{figure}

\begin{figure}[htb] \centering 
	\includegraphics[width=8cm]{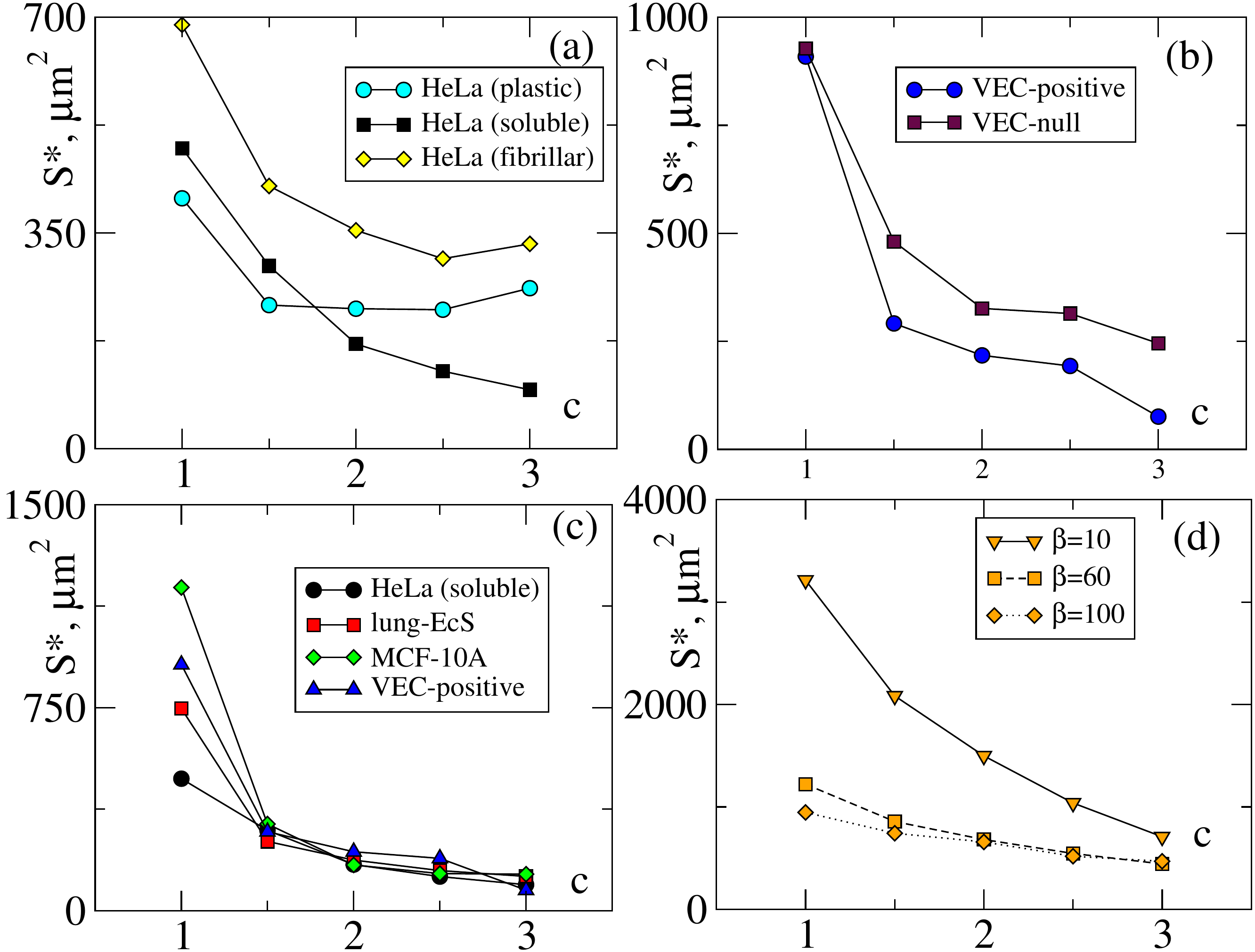}
	\caption{\label{fig:S7} The characteristic cluster size $S^*= 2\langle S^2\rangle/\langle S \rangle$ as a function of the velocity threshold $c$. a) Hela Cells on different substrates. b) Mouse endothelial cells expressing VE-cadherin or not. c) Comparison between different cell lines d) Simulations of the model for different values of $\beta$.  }
\end{figure}

\begin{figure}[htb] \centering 
	\includegraphics[width=8cm]{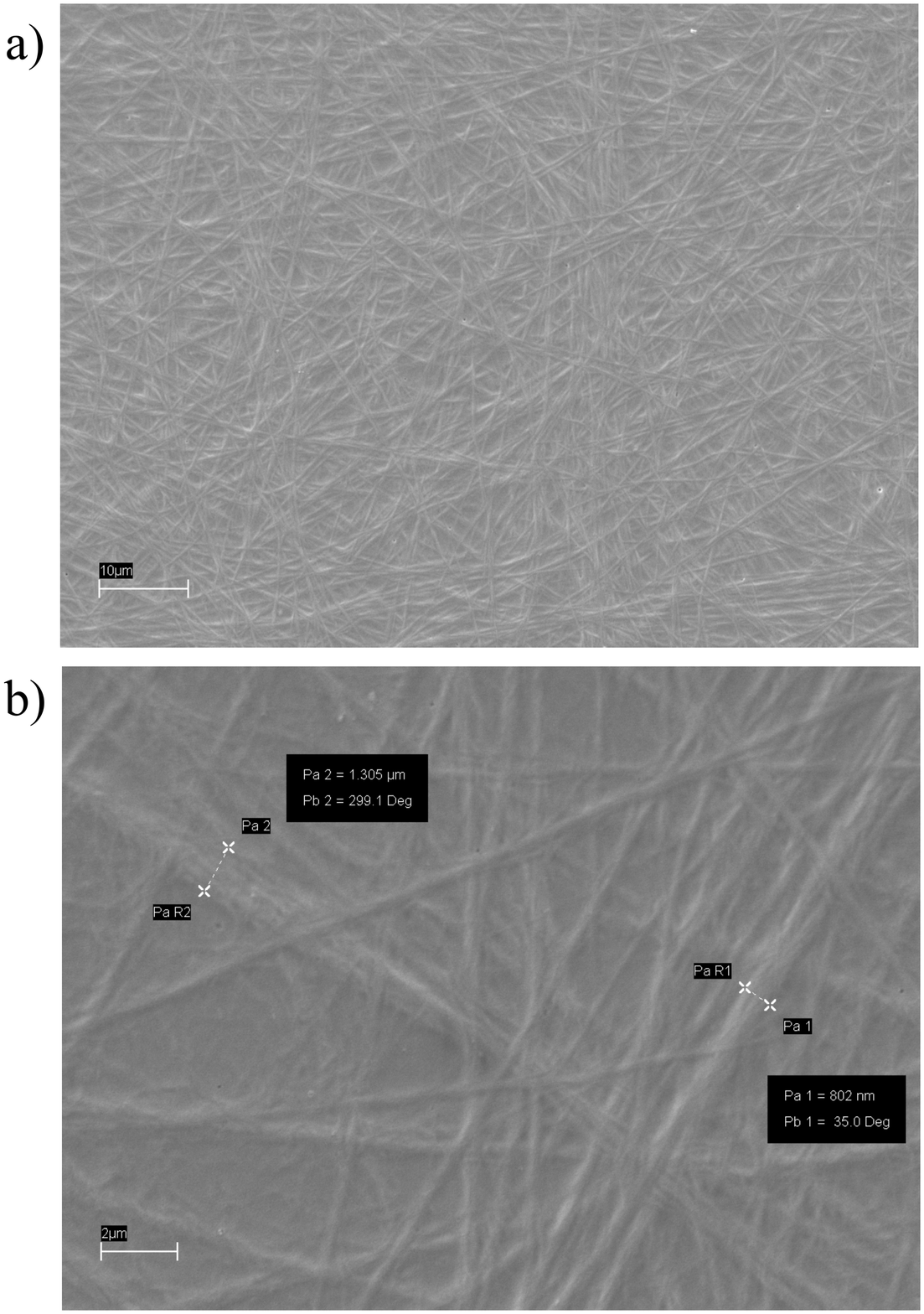}
	\caption{\label{fig:S3} Scanning electron micrograph of the fibrillar collagen substrate at two
		different magnifications. From the high magnification image b) it is possible to estimate the
		fiber diameter.}
\end{figure}

\begin{figure}[htb] \centering 
\includegraphics[width=8cm]{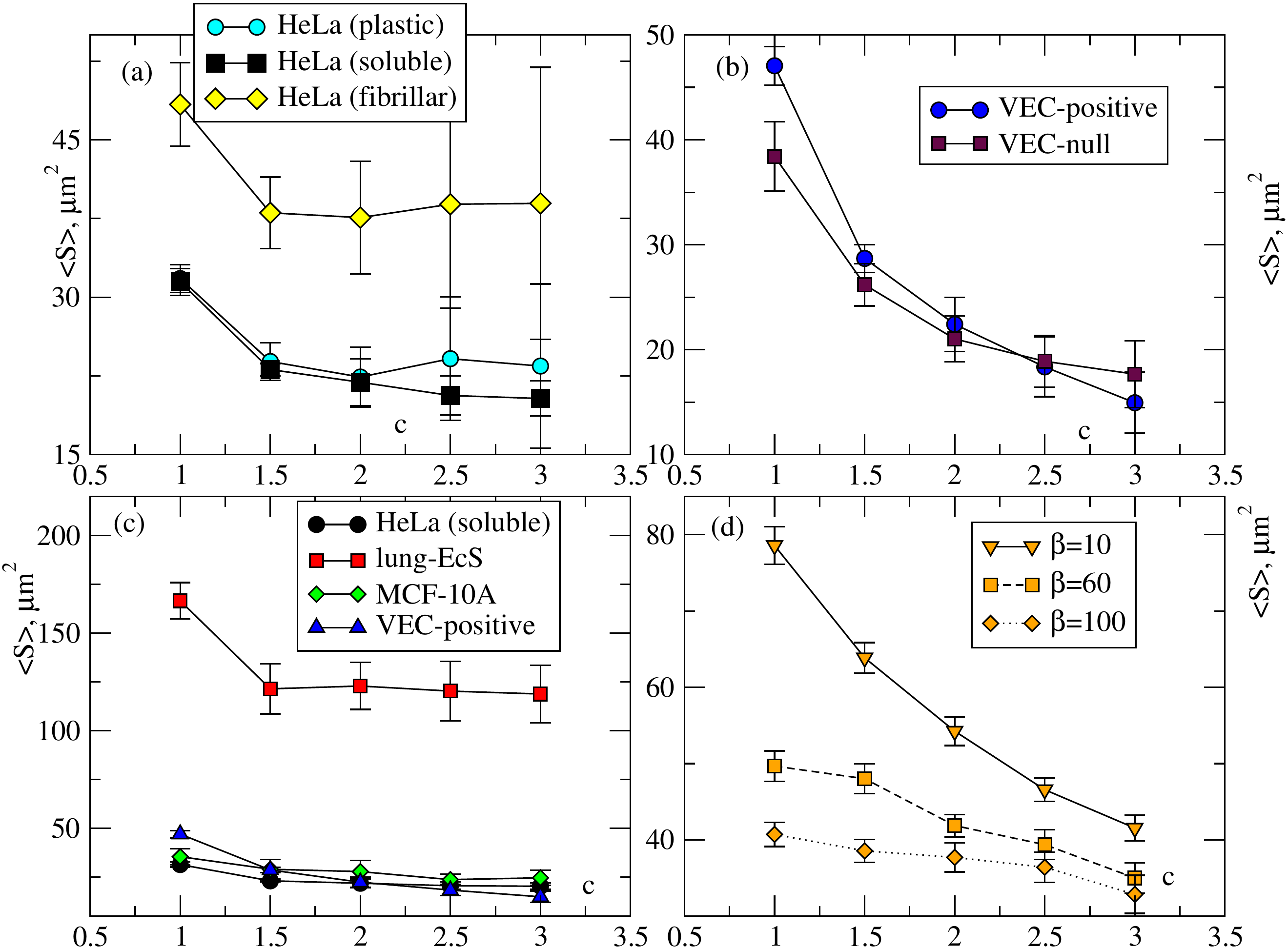}
 \caption{\label{fig:S2} The average cluster size as a function of the velocity threshold $c$. a) HeLa Cells
on different substrates. b) Mouse endothelial cells expressing VE-cadherin or not. c) Comparison between different
cell lines d) Simulations of the model for different values of $\beta$.  }
 \end{figure}

\begin{figure}[htb] \centering 
\includegraphics[width=8cm]{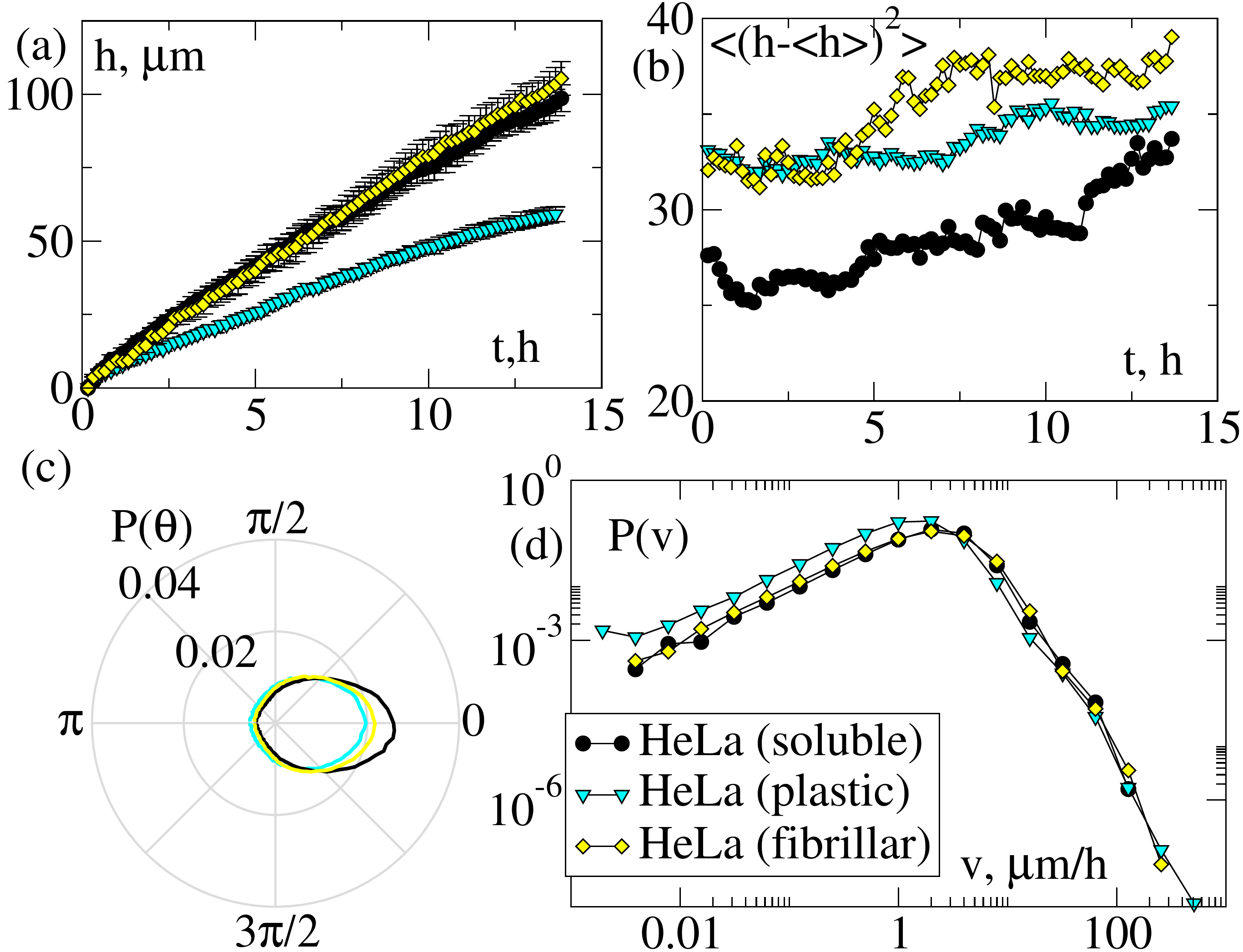}
 \caption{\label{fig:S4} Dependence of the front dynamics on the substrate properties. a) Front position as
 a function of time indicates that fronts move slowly on plastic substrates. b) The standard deviation of the 
 front (i.e. the front roughness) is larger for cells moving on fibrillar collagen. c) Angular velocity distribution
 as obtained from PIV.  d) Distribution of absolute values of velocities as obtained by PIV.}
 \end{figure}

\begin{figure}[htb] \centering 
\includegraphics[width=8cm]{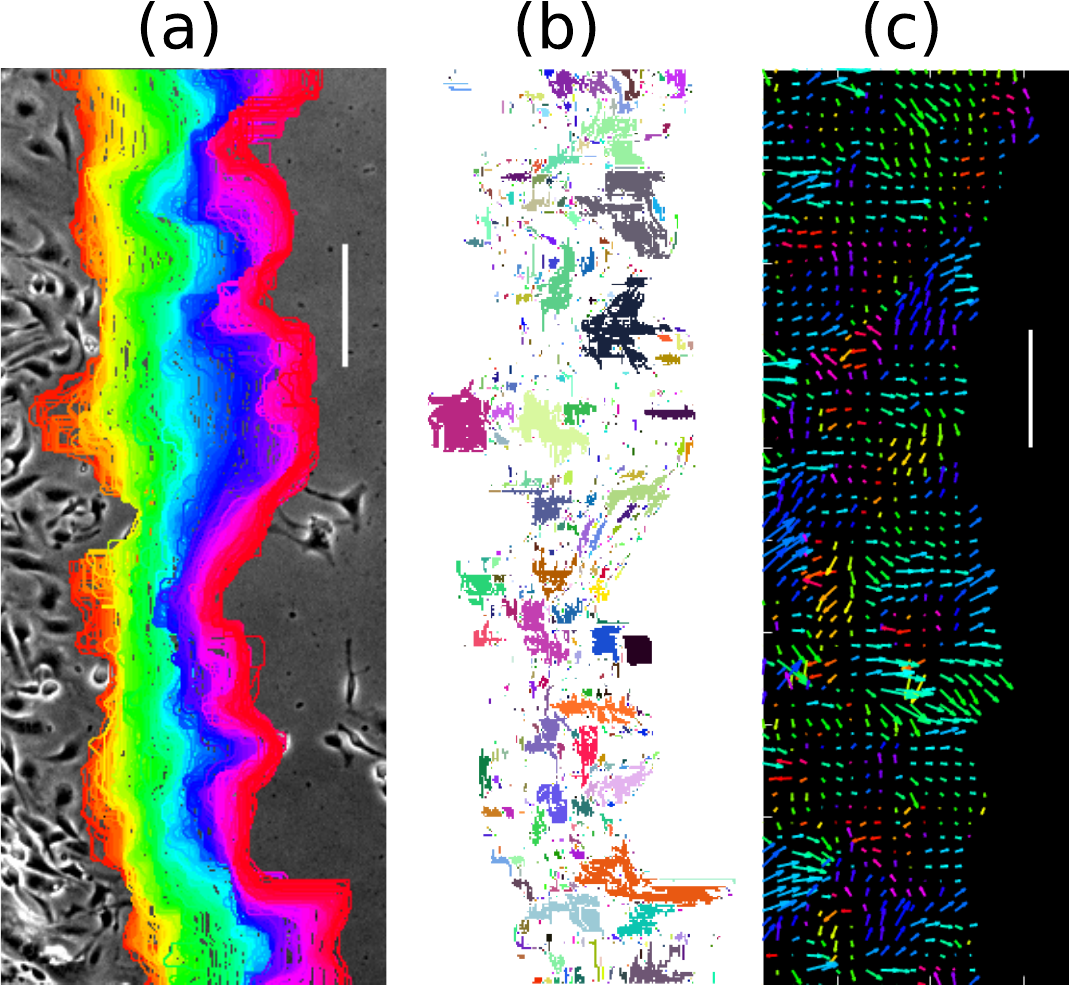}
 \caption{\label{fig:S5} Front dynamics in mouse endothelial cells with VE-cadherin knock down (VEC-null). a)
 Fronts. b) Activity clusters. c) Results from PIV. }
 \end{figure}

\begin{figure}[htb] \centering 
\includegraphics[width=8cm]{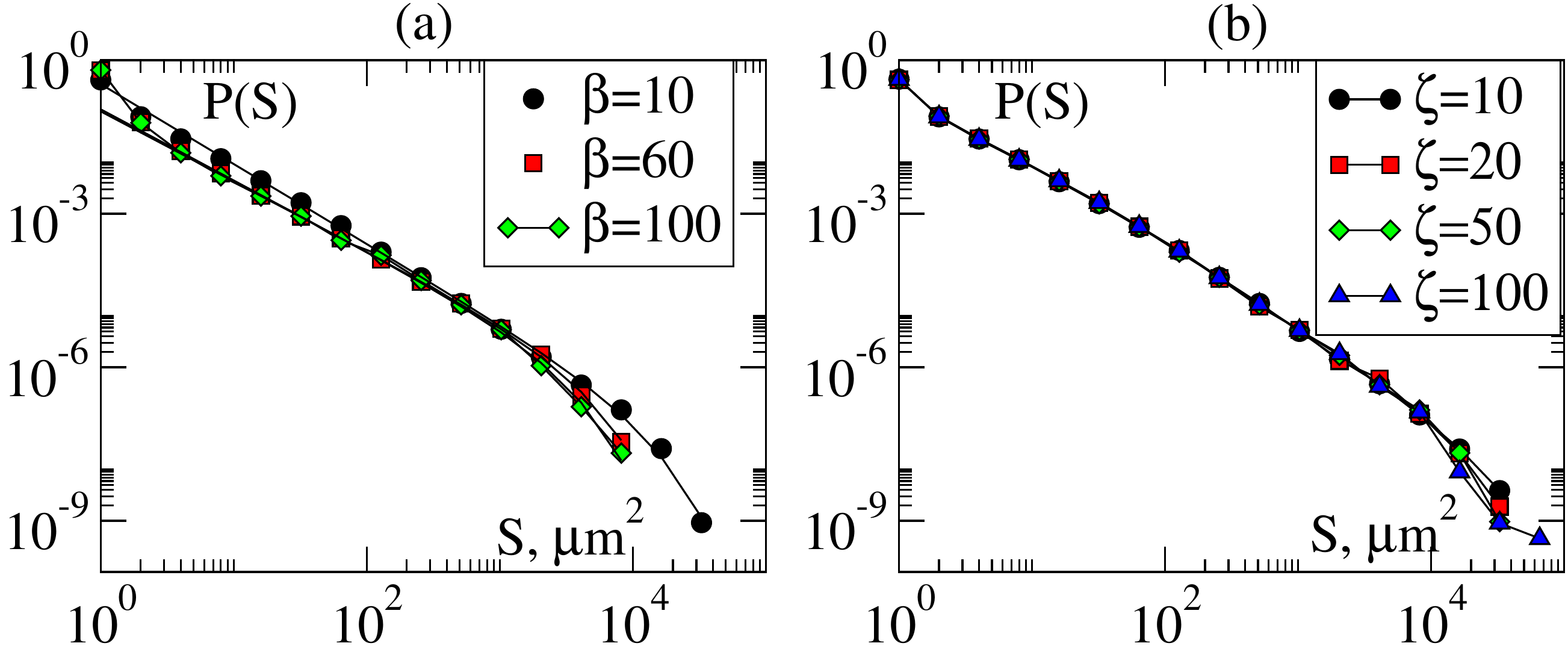}
 \caption{\label{fig:S6} Simulated cluster size distributions a) for different $\beta$ 
 (the strength of alignment between cells), b) for different $\zeta$ (correlation length of the random
 field). }
 \end{figure}

\begin{figure}[htb] \centering 
	\includegraphics[width=8cm]{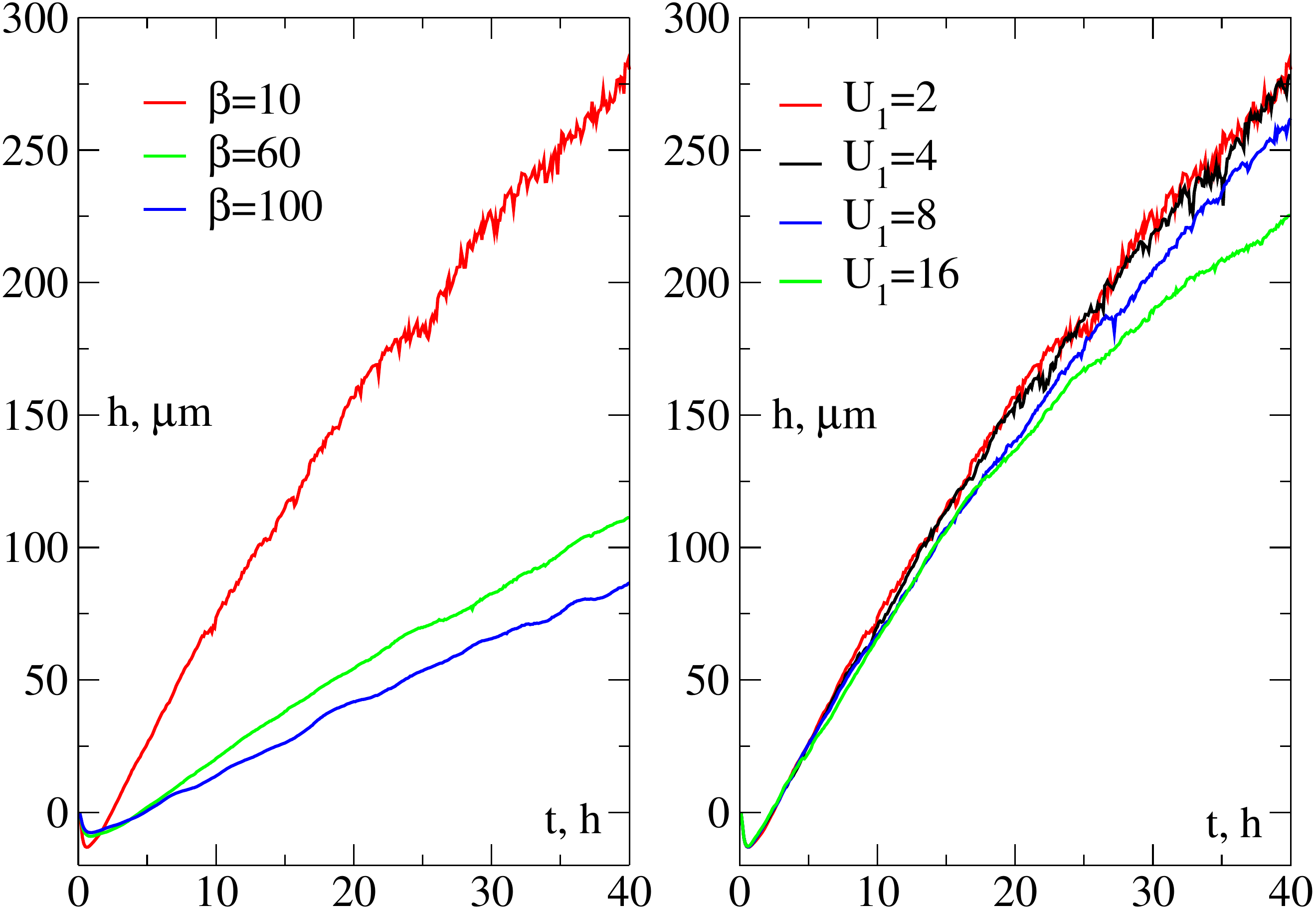}
	\caption{\label{fig:S8} a) Simulations of the model for different values of $\beta$ show that
		fronts are slower when $\beta$ is increased. b) Simulations of the model for different values of
		the adhesion strength $U_1$ show that
		fronts are slower when $U_1$ is increased.}
\end{figure}
 
\begin{figure}[htb] \centering 
	\includegraphics[width=8cm]{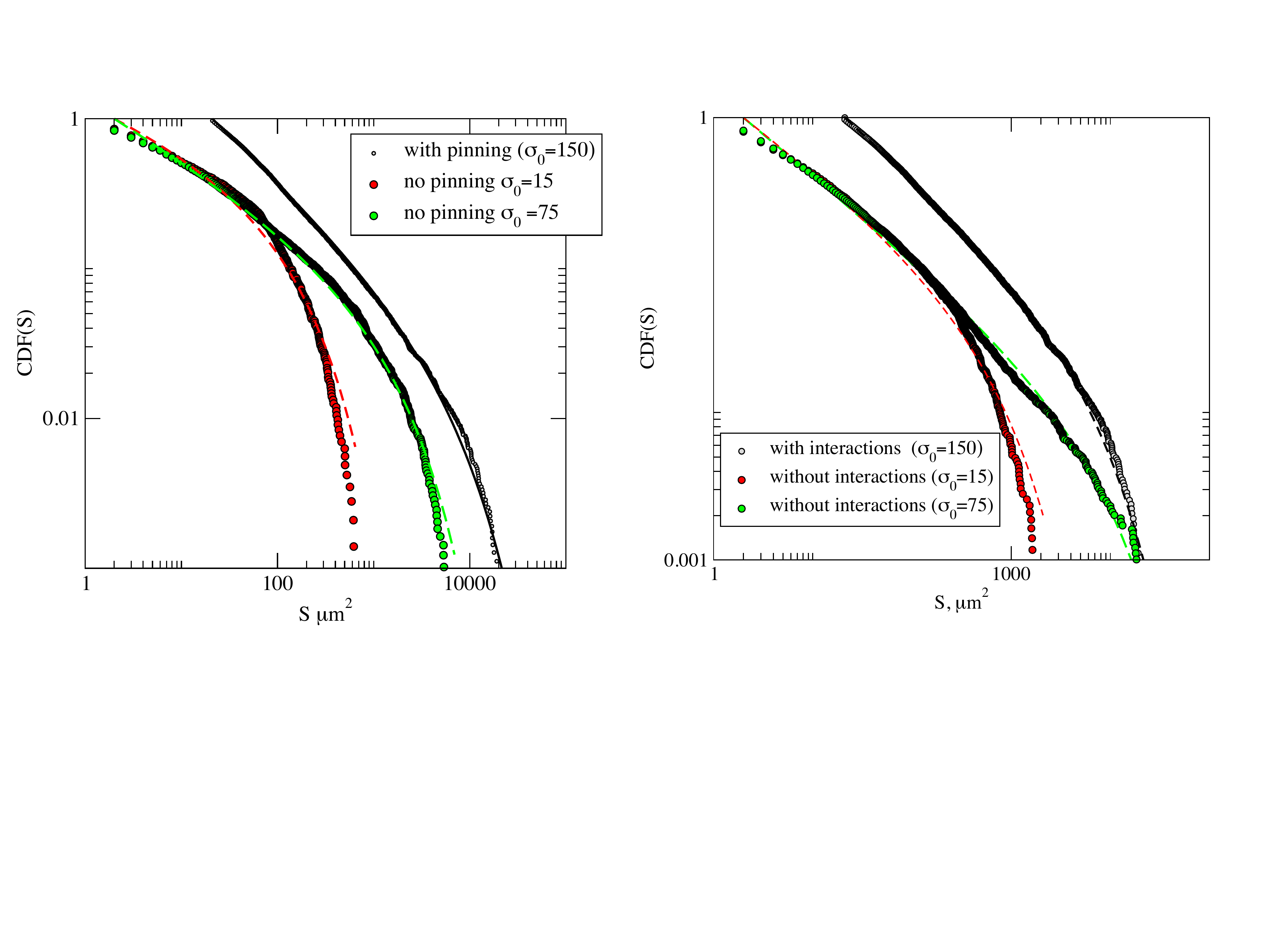}
	\caption{\label{fig:S9} The non-binned cumulative distribution functions (CDF) of the cluster sizes from simulations
		a) with and without pinning (i. e. no surface particles and no random field) and b) with and without interactions (i.e. $\beta=0$, $U_0=U_1=0$). The lack of either pinning or interactions leads to clear deviation from a power law distribution, as also shown by fitting the simulation results with the maximum likelihood method. The fit yield scaling exponents $\tau=1.2$ (panel a, red), $\tau=1.38$ (panel a, green), $\tau=1.51$ (panel b, red), $\tau=1.56$ (panel b, green), but
		the fitted curves deviate systematically from the data and the scaling regime is extremely small. 
		We also notice that in order to record a front at all in these conditions, one should also reduce the noise parameter $\sigma_0$.}
\end{figure}
 
\begin{figure}[htb] \centering 
	\includegraphics[width=14cm]{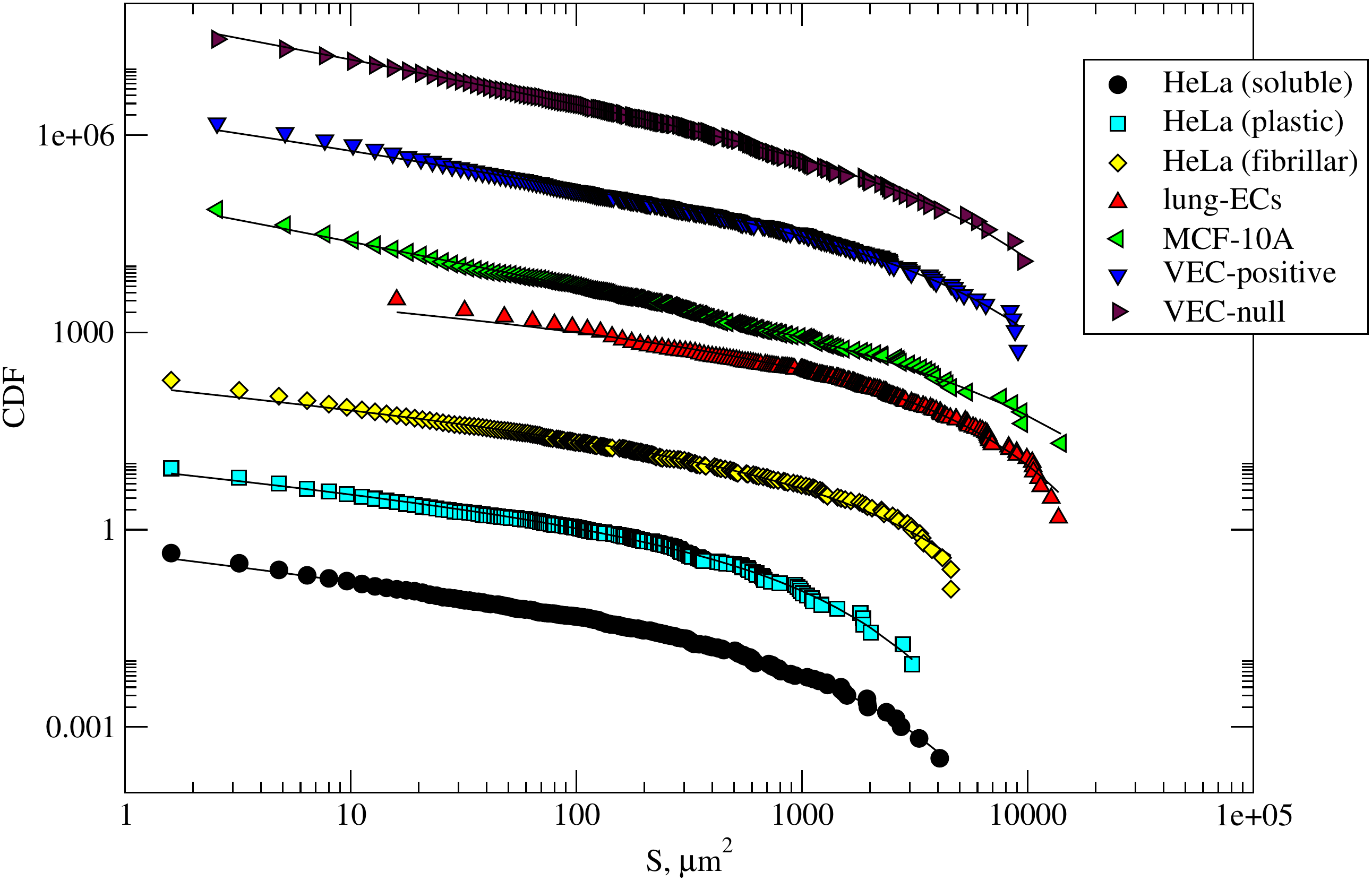}
	\caption{\label{fig:S10} The non-binned cumulative distribution functions (CDF) of the cluster sizes for the different
		experiments are compared with the results of maximum likelihood estimates.}
\end{figure}


\clearpage

\setcounter{table}{0}
\renewcommand{\thetable}{S\arabic{table}} 

\begin{center}
	\begin{table}[h]
		\begin{tabulary}{1.0\textwidth}{L| R R R | R R R |} 
			& \multicolumn{3}{|c|}{Fitting} &
			\multicolumn{3}{|c|}{Maximum likelihood} \\
			\hline
			Name & $A$ & $\tau$ & $C$ & $\tau$ & $C$ & $x_{min}$ \\
			\hline
			HeLa (soluble) & $0.18 \pm 0.02$ & $1.48 \pm 0.03$ & $0.00054 \pm 5.2 \times 10^{-5}$ &
			$1.48\pm 0.10$ & $(3.2 \pm 2.43) \times 10^{-4}$ & 12.8 \\ 
			HeLa (plastic) & $0.17 \pm 0.03$ & $1.47 \pm 0.05$ & $0.0006 \pm 0.0001$ & 
			$1.49 \pm 0.09$ & $(3.5 \pm 2.8) \times 10^{-4}$ & 8.0 \\
			HeLa (fibrillar) & $0.19 \pm 0.03$ & $1.49 \pm 0.04$  & $(20 \pm 7)\times 10^{-5}$ & 
			$1.34 \pm 0.14$ & $(2.82 \pm 2.0) \times 10 ^{-4}$  & 28.8 \\
			lung-ECs & $0.81 \pm 0.36$ & $1.58 \pm 0.07$ & $(7.2 \pm 1.4) \times 10^{-5}$
			& $1.45 \pm 0.13$ & $(1.1 \pm 0.6) \times 10^{-4}$ & 176 \\
			MCF-10A & $0.17 \pm 0.06$ & $1.58 \pm 0.08$ & $(16.7 \pm 9.2) \times 10^{-5}$
			& $1.56 \pm 0.12$ & $(1.2 \pm 1.0) \times 10^{-4}$ & 43.5 \\
			VEC-positive & $0.33 \pm 0.09$ & $1.62 \pm 0.05$ & $(8.6 \pm 4.9)\times 10^{-5}$ & 
			$1.55 \pm 0.12$ & $(1.2 \pm 1.0)\times 10^{-4}$ & 53.8 \\
			VEC-null & $0.33 \pm 0.04$ & $1.62 \pm 0.03$ & $0.0001 \pm 5 \times 10^{-5}$ & $1.61 \pm 0.13$ & $(1.7 \pm 1.4) \times 10^{-4}$ & 46.1 \\
			\hline
		\end{tabulary}
		\caption{Best parameters obtained for the cluster distributions for experimental data by least-square fitting
			of the probability density function and by the maximum likelihood estimate. The smallest value of the cluster
			size ($x_{min}$) is also reported in the table. \label{table:S1}}
	\end{table}
	\end{center}
	
	\begin{center}
	\begin{table}[h]
		\begin{tabulary}{1.0\textwidth}{ L | R R R | R R R |}
			& \multicolumn{3}{|c|}{Fitting} &
			\multicolumn{3}{|c|}{Maximum likelihood} \\
			\hline
			$\beta$ & $A$ & $\tau$ & $C$ & $\tau$ & $C$ & $x_{min}$ \\
			\hline
			$10$ & $0.34 \pm 0.07$ & $1.56 \pm 0.04$ & $(10 \pm 1)\times 10^{-5}$ &
			$1.60\pm 0.02$ & $(6.3 \pm 1.9) \times 10^{-4}$ & $21$ \\ 
			$60$ & $0.103\pm 0.007 $ & $1.37 \pm 0.01$ & $(3\pm 0.13 )\times 10^{-4}$ & 
			$1.36 \pm 0.05$  & $(2.5 \pm 0.8) \times 10^{-4}$ & $16$ \\
			$100$ & $0.11 \pm 0.02$ & $1.39 \pm 0.04$  & $(3.8\pm 0.3) \times 10^{-4}$ & 
			$1.33 \pm 0.03$ & $(3.9\pm 0.7) \times 10^{-4}$ & $6$ \\
			\hline
		\end{tabulary}
		\caption{Best parameters obtained for the cluster distributions for simulated data by least-square fitting
			of the probability density function and by the maximum likelihood estimate. The smallest value of the cluster
			size ($x_{min}$) is also reported in the table. \label{table:S2}}
	\end{table}
	\end{center}

		\begin{center}
	\begin{table}[h]
		\centering
		\begin{tabular}{|c|c|c|c| }
			\hline
			Name & Number of fronts & 1 px in $\mu$m & 1 frame in min. \\
			\hline
			HeLa (plastic) & 6 &1.2658 & 10 \\
			HeLa (soluble) & 5 &1.2658 & 10 \\
			HeLa (fibrillar) & 8 &1.2658 & 10 \\
			lung-ECs & 11 &4 & 5  \\
			MCF-10A & 8 &1.6 & 5 \\
			VEC-positive & 4 & 1.6 & 2 \\
			VEC-null & 4& 1.6 & 2 \\
			\hline
		\end{tabular}
		\caption{Time-lapse parameters. We list the number of fronts analyzed for each case,
			The size of a pixel in each case and sizes and time step between two frames \label{table:S3}.}
		
	\end{table}
		\end{center}

	

	

  
\clearpage 
  
\section*{Supplementary video captions}  
 
\begin{itemize}
\item[Video S1]  (Left) Time lapse of a representative sheet of Hela cells invading a fibrillar collagen substrate. The reconstructed front is reported in red. (Right) Evolution of the corresponding local velocity map obtained by PIV.
\item[Video S2] (Left) Time lapse of a representative sheet of Hela cells invading a soluble collagen substrate. The reconstructed front is reported in red. (Right) Evolution of the corresponding local velocity map obtained by PIV.
\item[Video S3] (Left) Time lapse of a representative sheet of Hela cells invading a plastic substrate. The reconstructed front is reported in red. (Right) Evolution of the corresponding local velocity map obtained by PIV.
\item[Video S4] (Left) Time lapse of a representative sheet of lung derived endothelial cells. The reconstructed front is reported in red. (Right) Evolution of the corresponding local velocity map obtained by PIV.
\item[Video S5] (Left) Time lapse of a representative sheet of MCF10-A cells invading a plastic substrate. The reconstructed front is reported in red. (Right) Evolution of the corresponding local velocity map obtained by PIV.
\item[Video S6] (Left) Time lapse of a representative sheet of VEC-null mouse endothelial cells invading a plastic substrate. The reconstructed front is reported in red. (Right) Evolution of the corresponding local velocity map obtained by PIV.
\item[Video S7] (Left) Time lapse of a representative sheet of VEC-positive  mouse endothelial cells invading a plastic substrate. The reconstructed front is reported in red. (Right) Evolution of the corresponding local velocity map obtained by PIV.
\item[Video S8] (Left) Representative simulation results. The reconstructed front is reported in red. (Right) Evolution of the corresponding local velocity map obtained by PIV.
\item[Video S9] A representative trajectory (yellow) of an isolated VEC-null mouse endothelial cell.
\end{itemize}


\begin{thebibliography}{10}
	
	\bibitem{Tambe2011}
	Tambe DT, et~al. (2011) Collective cell guidance by cooperative intercellular
	forces.
	\newblock \emph{Nat Mater} 10:469--75.
	
	\bibitem{Brugues2014}
	Brugues A, et~al. (2014) Forces driving epithelial wound healing.
	\newblock \emph{Nat Phys} 10:683--690.
	
	\bibitem{Haeger2014}
	Haeger A, Krause M, Wolf K, Friedl P (2014) Cell jamming: collective invasion
	of mesenchymal tumor cells imposed by tissue confinement.
	\newblock \emph{Biochim Biophys Acta} 1840:2386--95.
	
	\bibitem{Lange2013}
	Lange JR, Fabry B (2013) Cell and tissue mechanics in cell migration.
	\newblock \emph{Exp Cell Res} 319:2418--23.
	
	\bibitem{koch2012}
	Koch TM, M{\"u}nster S, Bonakdar N, Butler JP, Fabry B (2012) 3d traction
	forces in cancer cell invasion.
	\newblock \emph{PLoS One} 7:e33476.
	
	\bibitem{Sacks2003}
	Sacks MS, Sun W (2003) Multiaxial mechanical behavior of biological materials.
	\newblock \emph{Annu Rev Biomed Eng} 5:251--84.
	
	\bibitem{Vedula2013}
	Vedula SRK, Ravasio A, Lim CT, Ladoux B (2013) Collective cell migration: a
	mechanistic perspective.
	\newblock \emph{Physiology (Bethesda)} 28:370--9.
	
	\bibitem{Szabo2006}
	Szab\'o B, et~al. (2006) Phase transition in the collective migration of tissue
	cells: Experiment and model.
	\newblock \emph{Phys Rev E} 74:061908.
	
	\bibitem{Poujade2007}
	Poujade M, et~al. (2007) Collective migration of an epithelial monolayer in
	response to a model wound.
	\newblock \emph{Proc Natl Acad Sci U S A} 104:15988--93.
	
	\bibitem{Sepulveda2013}
	Sep{\'u}lveda N, et~al. (2013) Collective cell motion in an epithelial sheet
	can be quantitatively described by a stochastic interacting particle model.
	\newblock \emph{PLoS Comput Biol} 9:e1002944.
	
	\bibitem{Haga2005}
	Haga H, Irahara C, Kobayashi R, Nakagaki T, Kawabata K (2005) Collective
	movement of epithelial cells on a collagen gel substrate.
	\newblock \emph{Biophys J} 88:2250--6.
	
	\bibitem{Ng2012}
	Ng MR, Besser A, Danuser G, Brugge JS (2012) Substrate stiffness regulates
	cadherin-dependent collective migration through myosin-ii contractility.
	\newblock \emph{J Cell Biol} 199:545--63.
	
	\bibitem{Saez2007}
	Saez A, Ghibaudo M, Buguin A, Silberzan P, Ladoux B (2007) Rigidity-driven
	growth and migration of epithelial cells on microstructured anisotropic
	substrates.
	\newblock \emph{Proc Natl Acad Sci U S A} 104:8281--6.
	
	\bibitem{Rottgermann2014}
	R{\"o}ttgermann PJF, Alberola AP, R{\"a}dler JO (2014) Cellular
	self-organization on micro-structured surfaces.
	\newblock \emph{Soft Matter} 10:2397--404.
	
	\bibitem{Oakes2009}
	Oakes PW, et~al. (2009) Neutrophil morphology and migration are affected by
	substrate elasticity.
	\newblock \emph{Blood} 114:1387--95.
	
	\bibitem{Metzner2015}
	Metzner C, et~al. (2015) Superstatistical analysis and modelling of
	heterogeneous random walks.
	\newblock \emph{Nat Commun} 6:7516.
	
	\bibitem{Serra-Picamal2012}
	Serra-Picamal X, et~al. (2012) Mechanical waves during tissue expansion.
	\newblock \emph{Nat Phys} 8:628--634.
	
	\bibitem{Banerjee2015}
	Banerjee S, Utuje KJC, Marchetti MC (2015) Propagating stress waves during
	epithelial expansion.
	\newblock \emph{Phys Rev Lett} 114:228101.
	
	\bibitem{maloy2006}
	Maloy KJ, Santucci S, Schmittbuhl J, Toussaint R (2006) Local waiting time
	fluctuations along a randomly pinned crack front.
	\newblock \emph{Phys Rev Lett} 96:045501.
	
	\bibitem{Tallakstad2011}
	Tallakstad KT, Toussaint R, Santucci S, Schmittbuhl J, Maloy KJ (2011) Local
	dynamics of a randomly pinned crack front during creep and forced
	propagation: an experimental study.
	\newblock \emph{Phys Rev E Stat Nonlin Soft Matter Phys} 83:046108.
	
	\bibitem{Clotet2014}
	Clotet X, Ort{\'\i}n J, Santucci S (2014) Disorder-induced capillary bursts
	control intermittency in slow imbibition.
	\newblock \emph{Phys Rev Lett} 113:074501.
	
	\bibitem{durin00}
	Durin G, Zapperi S (2000) Scaling exponents for {B}arkhausen avalanches in
	polycrystalline and amorphous ferromagnets.
	\newblock \emph{Phys Rev Lett} 84:4075--4078.
	
	\bibitem{leschhorn97}
	Leschhorn H, Nattermann T, Stepanow S, Tang LH (1997) Driven interface
	depinning in a disordered medium.
	\newblock \emph{Ann Physik} 6:1--34.
	
	\bibitem{rosso2009}
	Rosso A, Le~Doussal P, Wiese KJ (2009) Avalanche-size distribution at the
	depinning transition: A numerical test of the theory.
	\newblock \emph{Phys Rev B} 80:144204.
	
	\bibitem{narayan92}
	Narayan O, Fisher DS (1992) Critical behavior of sliding charge-density waves
	in 4- epsilon dimensions.
	\newblock \emph{Phys Rev B} 46:11520.
	
	\bibitem{chauve01}
	Chauve P, Doussal PL, Wiese KJ (2001) Renormalization of pinned elastic
	systems: how does it work beyond one loop.
	\newblock \emph{Phys Rev Lett} 86:1785--1788.
	
	\bibitem{ledoussal2009}
	Le~Doussal P, Wiese KJ (2009) Size distributions of shocks and static
	avalanches from the functional renormalization group.
	\newblock \emph{Phys Rev E} 79:051106.
	
	\bibitem{sethna01}
	Sethna J, Dahmen KA, Myers CR (2001) Crackling noise.
	\newblock \emph{Nature} 410:242--244.
	
	\bibitem{giampietro2015}
	Giampietro C, et~al. (2015) The actin-binding protein eps8 binds ve-cadherin
	and modulates yap localization and signaling.
	\newblock \emph{J Cell Biol} 211:1177--92.
	
	\bibitem{Goodwin2004}
	Goodwin M, Yap AS (2004) Classical cadherin adhesion molecules: coordinating
	cell adhesion, signaling and the cytoskeleton.
	\newblock \emph{J Mol Histol} 35:839--44.
	
	\bibitem{Wheelock2003}
	Wheelock MJ, Johnson KR (2003) Cadherins as modulators of cellular phenotype.
	\newblock \emph{Annu Rev Cell Dev Biol} 19:207--35.
	
	\bibitem{giampietro2012}
	Giampietro C, et~al. (2012) Overlapping and divergent signaling pathways of
	n-cadherin and ve-cadherin in endothelial cells.
	\newblock \emph{Blood} 119:2159--70.
	
	\bibitem{clauset2009}
	Clauset A, Shalizi CR, Newman ME (2009) Power-law distributions in empirical
	data.
	\newblock \emph{SIAM review} 51:661--703.
	
	\bibitem{Angelini2011}
	Angelini TE, et~al. (2011) Glass-like dynamics of collective cell migration.
	\newblock \emph{Proc Natl Acad Sci U S A} 108:4714--9.
	
	\bibitem{Park2015}
	Park JA, et~al. (2015) Unjamming and cell shape in the asthmatic airway
	epithelium.
	\newblock \emph{Nat Mater} 14:1040--1048.
	
	\bibitem{Weber2015}
	Weber CA, et~al. (2015) Random bursts determine dynamics of active filaments.
	\newblock \emph{Proceedings of the National Academy of Sciences}
	112:10703--10707.
	
	\bibitem{Ginelli2015}
	Ginelli F, et~al. (2015) Intermittent collective dynamics emerge from
	conflicting imperatives in sheep herds.
	\newblock \emph{Proc Natl Acad Sci U S A} 112:12729--34.
	
	\bibitem{Song2010}
	Song C, Koren T, Wang P, Barabasi AL (2010) Modelling the scaling properties of
	human mobility.
	\newblock \emph{Nat Phys} 6:818--823.
	
	\bibitem{balconi2000}
	Balconi G, Spagnuolo R, Dejana E (2000) Development of endothelial cell lines
	from embryonic stem cells: A tool for studying genetically manipulated
	endothelial cells in vitro.
	\newblock \emph{Arterioscler Thromb Vasc Biol} 20:1443--51.
	
	\bibitem{Lampugnani2002}
	Lampugnani MG, et~al. (2002) Ve-cadherin regulates endothelial actin activating
	rac and increasing membrane association of tiam.
	\newblock \emph{Mol Biol Cell} 13:1175--89.
	
	\bibitem{PIV0}
	Thielicke W (2014) \emph{The flapping flight of birds: Analysis and
		application}.
	\newblock Ph.D. thesis.
	
	\bibitem{PIV1}
	Thielicke W, Stamhuis E (2014) Pivlab – towards user-friendly, affordable and
	accurate digital particle image velocimetry in matlab 2.
	
	\bibitem{Johnston2014}
	Johnston ST, Simpson MJ, McElwain DLS (2014) How much information can be
	obtained from tracking the position of the leading edge in a scratch assay?
	\newblock \emph{Journal of The Royal Society Interface} 11.
	
	\bibitem{Treloar2013}
	Treloar KK, Simpson MJ (2013) Sensitivity of edge detection methods for
	quantifying cell migration assays.
	\newblock \emph{PLoS One} 8:e67389.
	
\end{thebibliography}

%

\end{article}
 \end{document}